\documentclass[aps,prc,twocolumn,superscriptaddress]{revtex4-1}

\usepackage{longtable}
\usepackage{grffile} %deal with fig name with ".,_ " .etc...
\usepackage{graphics}
\usepackage{graphicx} %set figure size with width=...
\usepackage{amsmath}    % need for subequations
\usepackage{hyperref}   % use for hypertext links, including those to external documents and URLs
\usepackage{epsfig}
\usepackage{epstopdf}
\usepackage{color}
\usepackage{url}

\usepackage{lineno}
\begin{document}
\title{Improvement of heavy flavor productions in a multi-phase
  transport model updated with modern nPDFs}
\author{L. Zheng}\email{zhengliang@cug.edu.cn}
\affiliation{School of Mathematics and Physics, China University of
  Geosciences (Wuhan), Wuhan 430074, China}
\affiliation{Key Laboratory of Quark and Lepton Physics (MOE) and Institute
of Particle Physics, Central China Normal University, Wuhan 430079, China}
\author{C. Zhang}
\affiliation{Key Laboratory of Quark and Lepton Physics (MOE) and Institute
of Particle Physics, Central China Normal University, Wuhan 430079, China}
\author{S.S. Shi} \email{shiss@mail.ccnu.edu.cn}
\affiliation{Key Laboratory of Quark and Lepton Physics (MOE) and Institute
of Particle Physics, Central China Normal University, Wuhan 430079, China}
\author{Z.W. Lin}\email{linz@ecu.edu}
\affiliation{Key Laboratory of Quark and Lepton Physics (MOE) and Institute
of Particle Physics, Central China Normal University, Wuhan 430079, China}
\affiliation{Department of Physics, East Carolina University,
  Greenville, North Carolina 27858, USA} 

%\date{\today}

\begin{abstract}
Recently we have updated a multi-phase transport (AMPT) model with
modern parton distribution functions of nuclei (nPDFs). 
Here we study open charm production in the updated AMPT model 
and compare to the experimental data from $pp$ and $AA$ collisions
over a wide range of collision energies.
Besides the update of nPDFs, we have removed the transverse momentum
cutoff on initial heavy quark productions and also included the
resultant heavy flavor cross section into the total minijet 
cross section in the initial condition as described by the HIJING model. 
We show that the AMPT model with these updates provides a much better
description of the yields and transverse momentum spectra
of various open charm hadrons in comparison with the  experimental
data. This lays the foundation for further heavy flavor studies within the
transport model approach. 
\end{abstract}

\maketitle

\section{Introduction}
\label{sec:intro}
In high energy hadronic collisions, heavy flavor production provides 
us a powerful tool to study quantum
chromodynamics (QCD)~\cite{Andronic:2015wma}. The initial production 
of heavy flavor quarks is calculable
with the perturbative QCD (pQCD) 
method due to the relatively large value of the heavy quark mass. 
In heavy ion physics, heavy quarks also play an important role because
their masses are typically larger than the temperatures achieved in
the produced quark-gluon plasma (QGP). Therefore, they are
predominantly produced in the initial hard scatterings between the two
nuclei on a time scale shorter than the formation time of the QGP. 
As a result, heavy quarks can experience almost the full
evolution of the deconfined nuclear medium and are 
thus sensitive to early dynamics of the collision system 
~\cite{Muller:1992xn,Lin:1994xma}.

Recently, it has been realized that the
strong electromagnetic fields in the initial state of heavy ion
collisions may significantly affect the heavy quark directed flow
$v_1$ and result in a larger charm $v_1$ than lighter
particles~\cite{Das:2016cwd,Chatterjee:2017ahy,Nasim:2018hyw}. 
In addition, heavy quarks or hadrons are also expected to interact
with the QGP or hadronic medium through elastic or inelastic processes  
during their propagation in the dense matter. 
This would lead to the suppression of heavy hadron yields at high
transverse momentum often represented by the nuclear modification
factor $R_{AA}$ and anisotropic flows of heavy hadrons such as
elliptic flow $v_2$. These observables can be used to extract the
transport properties of the QGP matter such as the drag and diffusion 
coefficients~\cite{He:2012df,Huggins:2012dj,Lang:2012cx}. 
For example, a large suppression in $R_{AA}$ and/or a substantially
non-zero $v_2$ for open heavy particles indicates that heavy quarks
experience significant interactions with the bulk medium.

Multiple theoretical frameworks have been developed to describe heavy
flavor productions in high energy collisions. 
The fixed flavor number scheme (FFNS)~\cite{Mangano:1991jk} is the
simplest scheme for the treatment of heavy flavors in the pQCD theory. 
Next-to-leading order (NLO) calculations are available for this approach,
while the gluon fragmentation to heavy flavor hadrons is not included.
Other implementations have been developed on the basis of the FFNS method.
Results from the general-mass variable-flavor-number scheme (GM-VFNS) 
approach~\cite{Kniehl:2004fy,Helenius:2018uul} as an extension to FFNS 
generally agree well with both the $pp$ and $p$A data in a wide
rapidity range. The fixed order plus next-to-leading logarithms (FONLL)
formalism~\cite{Cacciari:2005rk} is another widely used pQCD method 
that matches the massive FFNS cross section with the massless VFNS.
This approach usually can reasonably describe the open charm experimental
data, whereas its central value often under-predicts the data. 

In addition, medium induced effects can be included with models
based on the heavy quark transport or pQCD calculations of the parton
energy loss. At the high transverse 
momentum ($p_{\rm T}$) region, models~\cite{Djordjevic:2015hra,Xu:2015bbz}
that include both collisional and radiative energy loss of heavy quarks usually
provide a fair description of the $R_{AA}$ from central to peripheral
collisions. 
On the other hand, the evolution of low-$p_{\rm T}$ heavy quarks in 
the bulk medium is similar to the Brownian motion and can thus be studied with
transport approaches based on the Langevin or 
Boltzmann equation, which has been implemented in the models like
POWLANG~\cite{Beraudo:2014boa}, TAMU~\cite{He:2014cla},
Duke~\cite{Cao:2015hia}, BAMPS~\cite{Uphoff:2014hza},
LBT~\cite{Cao:2017hhk}, MC$@$HQ+EPOS~\cite{Nahrgang:2013xaa},
PHSD~\cite{Song:2015ykw} and
Catania~\cite{Das:2015ana,Plumari:2017ntm}.

A multi-phase transport (AMPT) model~\cite{Lin:2004en} is a useful
tool to study the bulk medium through the microscopic dynamical 
processes of the evolving system.  
The ZPC component solves the Boltzmann equation for two-body
scatterings via the parton cascade approach~\cite{Zhang:1997ej}.
As a self-contained event generator, the AMPT model provides a unified framework to
explore the medium evolution with different flavors including the 
event-by-event fluctuation and conservation of conserved charges. 
For example, the initial production of heavy quarks 
is modeled together with that of the light quarks and therefore 
the conservation of quantities such as energy, momentum, net
baryon number, and net charm number is guaranteed in the AMPT initial
condition of each event. 
Transport model studies of the dense matter, 
including heavy flavor studies with the AMPT
model~\cite{Zhang:2005ni,Li:2018leh}, 
help us to understand the QGP evolution and the transition from 
the non-equilibration stage to the hydrodynamic stage
~\cite{He:2015hfa,Lin:2015ucn,Kurkela:2018wud,Kurkela:2018vqr}. 
In particular, systematic comparisons of model predictions
with the measured heavy flavor $R_{AA}$ and
$v_2$~\cite{Cao:2018ews,Xu:2018gux} allow us to determine   
the relevant QGP medium transport coefficients such as the spatial
diffusion constant $D_{s}$, the drag coefficient $\eta_{D}$, and the
momentum transport coefficients $\kappa_{L},\kappa_{T},\hat{q}$.

Recently we have updated the AMPT model with an improved quark
coalescence process~\cite{He:2017tla} and modern nuclear parton
distribution functions (nPDFs)~\cite{Zhang:2019utb}. 
In this work, we will use the improved AMPT model to address open
charm production and then compare the model results with the
experimental data. The rest of the paper is organized as follows. In
Sec.~\ref{sec:hf_ampt} we discuss the physics changes we have made
for open heavy flavor productions in the AMPT model. 
We then calculate the yield and $p_{\rm T}$ spectra of various open
charm hadrons in comparison with the experimental data in $pp$ and $AA$
collisions in Sec.~\ref{sec:charm_hadron_pp}
and Sec.~\ref{sec:charm_hadron_AA}, respectively. 
After discussions in Sec.~\ref{sec:discussions}, we summarize
in Sec.~\ref{sec:summary}. 

\section{Descriptions of heavy flavors in the updated AMPT model}
\label{sec:hf_ampt}

In the AMPT model, the initial production of heavy quarks ($Q$) is 
handled by the HIJING two-component model~\cite{Gyulassy:1994ew}. 
It includes pair productions ($q+\bar{q} \rightarrow 
Q+\bar{Q}$, $g+g \rightarrow Q+\bar{Q}$) and gluon splitting 
($g \rightarrow Q+\bar{Q}$). 
The gluon splitting process is implemented with the parton shower method
similar to that in general purpose Monte Carlo event generators ~\cite{Sjostrand:2006za,Bahr:2008pv},
which includes the leading log resummation of the multiple parton 
emission phenomenon~\cite{Norrbin:2000zc}. 
However, the flavor excitation processes 
($q+Q \rightarrow q+Q$, $g+Q \rightarrow g+Q$) are not included. 
This is partly because the HIJING model only deals with closed
string objects created by minijet productions
but a flavor excitation process usually delivers a single heavy quark
jet in the final state.  
The pair production cross section for heavy quarks in pQCD at leading
order can be expressed as 
\begin{eqnarray}
\frac {d\sigma^{Q\bar{Q}}} {dp_{\rm T}^{2} dy_{1} dy_{2}}
\!\!= \!\! K  \! \sum_{a,b} \!\! 
x_{1}f_{a}(x_1,\!\mu_{F}^{2}\!)x_{2}f_{b}(x_2,\! \mu_{F}^{2}\!)  \!   
\frac {d\sigma^{ab \rightarrow  Q\bar{Q}}} {d\hat {t}}.
\label{eqn:dsigQQ}
\end{eqnarray}
In the above, 
$y_{1}$ and $y_{2}$ are respectively the rapidity of the two produced
partons, the $K$ factor aims to account for higher-order
corrections of heavy quark productions, $a$ and $b$ refer to the type
of interacting partons in the initial state, $x$ denotes the nucleon
momentum fraction taken by the interacting parton, $\mu_{F}$
represents the factorization scale, $f_a$ represents 
the parton distribution function of parton type $a$ in a (free or
bound) nucleon, and $\sigma^{ab \rightarrow Q\bar{Q}}$ is the cross
section for parton types $a$ and $b$ to produce the heavy quark pair. 
The $K$-factor $K=2.5$ is used for both light and heavy
flavor productions, as done in our previous study of light flavor
observables after the update of nPDFs~\cite{Zhang:2019utb}. 
For the charm quark production we take
$\mu_{F}^2=2(p^2_{\perp}+m_{c}^{2})$, where $p_{\perp}$ is the
transverse momentum transfer and we take the charm
quark mass as $m_{c}=1.3$ GeV$/c^2$ in this study.

The calculation of hard scatterings in AMPT is implemented within the
HIJING two-component model, where each hard parton is generated with a minimum
transverse momentum cut, $p_0$, to regulate the total minijet
production cross section. The differential minijet cross section
at leading order has the same form as Eq.(\ref{eqn:dsigQQ}):
\begin{eqnarray}
\frac {d\sigma^{cd}} {dp_{\rm T}^{2}dy_{1}dy_{2}} 
\!\!= \!\! K  \! \sum_{a,b} 
x_{1}f_{a}(x_1,\!\mu_{F}^{2}\!)x_{2}f_{b}(x_2,\! \mu_{F}^{2}\!)  \!   
\frac {d\sigma^{ab \rightarrow cd}} {d\hat {t}},
\label{eqn:dsigjet}
\end{eqnarray}
where $\sigma^{ab \rightarrow cd}$ is the cross section for parton
types $a$ and $b$ to produce a pair of minijets $c$ and $d$.
Then the total minijet cross section can be written as 
\begin{eqnarray}
\sigma_{\mathrm  {jet}}= \sum_{c,d} \frac {1} {1+\delta_{cd}} 
\int_{p_0^2}^{s/4} dp_{\rm  T}^2dy_{1}dy_{2}
\frac {d\sigma^{cd}} {dp_{\rm T}^2dy_{1}dy_{2}}.
\label{eqn:sigjet}
\end{eqnarray}

This minijet transverse momentum cutoff $p_0$ and the soft interaction
cross section $\sigma_{\rm soft}$ are the two key parameters in the HIJING 
two-component model, which control the elastic, inelastic and total
cross sections of $pp$ and $p\bar p$ collisions
\cite{Wang:1991hta,Gyulassy:1994ew,Deng:2010mv,Zhang:2019utb}.
Note that there is an extra factor of $1/2$ for final states with identical 
partons, such as $g+g \rightarrow g+g$ for minijet gluon productions,
in the above equation. In contrast, the original HIJING model applies
the factor of $1/2$  to all light flavor minijet production processes 
\cite{Wang:1991hta}, and that leads to a slightly smaller 
total $\sigma_{\rm jet}$ than Eq.(\ref{eqn:sigjet}) (at the same $p_0$).

In the HIJING model as well as the previous AMPT model (denoted as
``old AMPT'') before our most recent updates as done in
Ref.~\cite{Zhang:2019utb} and this study, 
the minijet cross section of Eq.(\ref{eqn:sigjet}) 
does not include the cross section of heavy flavors such as charm and
bottom quarks as given by Eq.(\ref{eqn:dsigQQ}). 
As a result, $\sigma_{\mathrm  {jet}}$, which is used 
in the eikonal formalism for the total, elastic and inelastic cross
sections~\cite{Wang:1991hta,Gyulassy:1994ew}, 
represents the cross section of light flavor ($u/d/s$) minijets. 
Then in the actual generation of minijets, 
heavy flavor minijets are still being generated and their fraction is 
calculated by the ratio of the heavy flavor cross section over the
total minijet cross section that includes both light and heavy
flavors, where the same minimum transverse momentum cut $p_0$ is used
in calculating $\sigma^{Q\bar{Q}}$ as done in Eq.(\ref{eqn:sigjet})
for $\sigma_{\rm jet}$.

The above approach has two issues. 
First, for self consistency the heavy flavor cross sections 
need to be included in the total minijet cross section 
in the two-component model. 
Secondly, the heavy quark cross sections can be calculated with pQCD
without any minimum transverse momentum requirement and 
the large heavy quark mass (compared to $\Lambda_{\rm QCD}$) 
naturally regulates the heavy quark total cross section. 
Therefore, we make significant changes to the 
descriptions of heavy flavors in the AMPT model by 
removing the $p_0$ cut for the heavy quark production cross sections 
and then including them in the total minijet cross section. 
These changes can be illustrated by the following 
modified formula for the new minijet cross section:
\begin{eqnarray}
\sigma_{\mathrm  {jet}}=&& \!\!\!\! \sum_{c,d} \frac {1} {1+\delta_{cd}} 
\int_{p_0^2}^{s/4} dp_{\rm T}^2dy_{1}dy_{2}
\frac {d\sigma^{cd}_{\mathrm  {light}}} {dp_{\rm T}^2dy_{1}dy_{2}}
\\ \nonumber
&+&\sum_{c,d}\int_{0}^{s/4}  dp_{\rm T}^2dy_{1}dy_{2}
\frac {d\sigma^{cd}_{\mathrm  {heavy}}} {dp_{\rm T}^2dy_{1}dy_{2}},
\label{eqn:sigjet_reform}
\end{eqnarray}
where the first term represents the total cross section of light
flavor ($u/d/s/g$) minijets and the second term represents that of
heavy flavor minijets including charm and bottom flavors. 

Naively one may expect that including the heavy quark production cross sections 
in the total minijet cross section will have negligible effects
because heavy quarks are very rare. However, in the two-component
model such as HIJING the light flavor ($u/d/s/g$) minijets require a
minimum $p_0$ while we think the heavy flavor minijets should not. 
Therefore when the $p_0$ value is high, which is
especially the case for both HIJING2.0 \cite{Deng:2010mv} and the
updated AMPT model~\cite{Zhang:2019utb} that use newer PDFs, 
the charm production cross section may be a significant fraction of
the total minijet cross section for $AA$ collisions at high energies.
Note that in the AMPT model updated with new nPDFs 
we have related the $p_0$ value for central $AA$ collisions at high
energies to that for $pp$ collisions with a nuclear
scaling of $p_0$~\cite{Zhang:2019utb}.  
That leads to a larger $p_0$ in $AA$ collisions and thus suppresses
light flavor minijet productions at high energies, while the initial heavy
flavor production is not affected.

\begin{figure}
\includegraphics[width=0.54\textwidth]{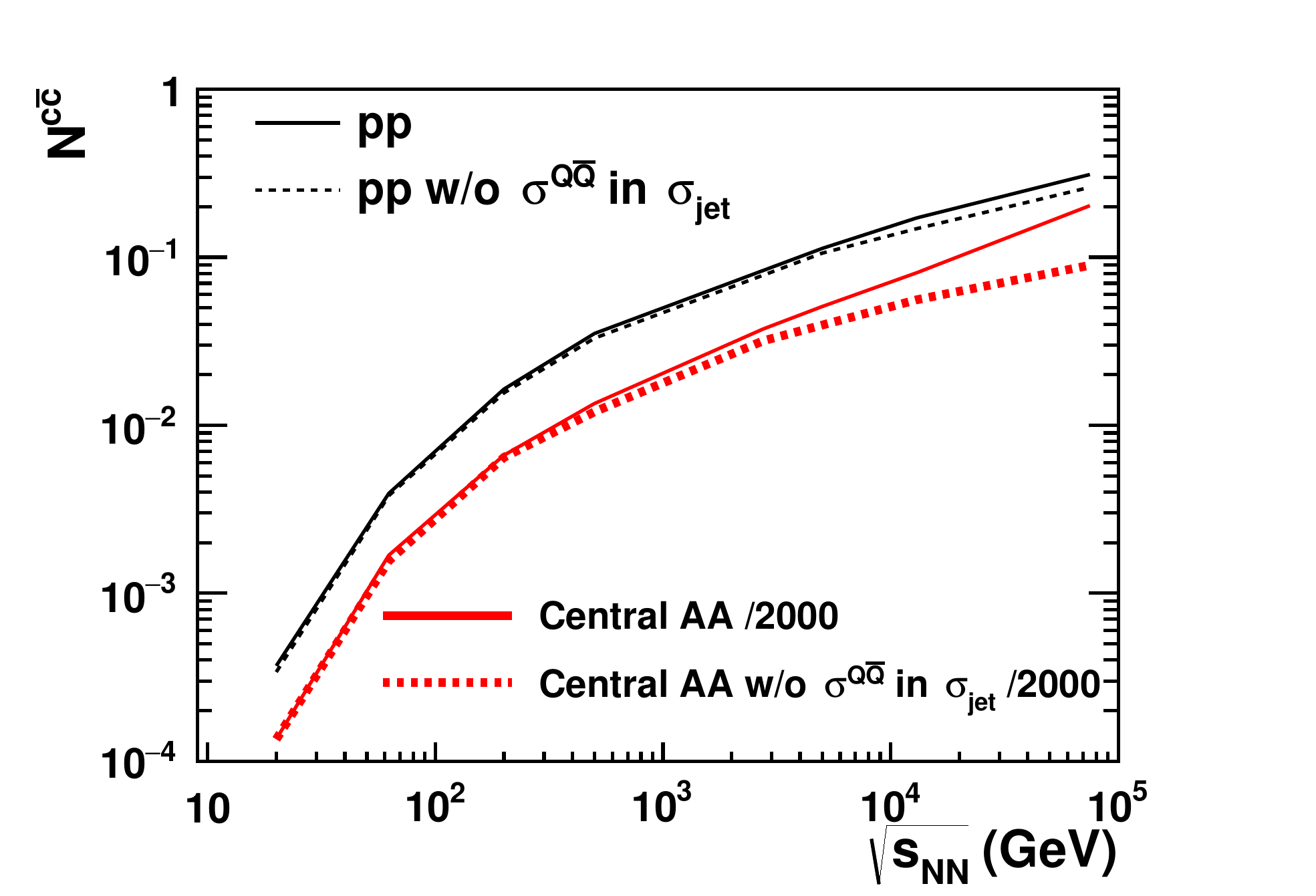}
\caption{(Color online) Effect of including heavy flavor cross
  sections into the minijet   cross section on the yield of
  charm-anticharm quark pairs from   the AMPT model for $pp$ and
  central $AA$ collisions (Au+Au at or below 200 GeV and Pb+Pb above
  200 GeV). 
  \label{fig:Nccbar_energy}
}
\end{figure}

The event averaged $c\bar{c}$ yield as a function of the collision energy 
is shown in Fig.~\ref{fig:Nccbar_energy}. Results without heavy quark channel in
the total minijet cross section are represented by the dashed curves while the
solid curves show the charm quark pair numbers after heavy flavor channels are included
as done in Eq.(~\ref{eqn:sigjet_reform}). The inclusion of heavy quark cross section
slightly increases the charm quark yield at high energies in $pp$ collisions. At low energies, the difference between
these two curves become negligible as expected due to the small heavy
flavor cross section. In $AA$ collisions, the impact of including the heavy flavor
cross section in $\sigma_{\rm jet}$ becomes more significant, up to a
factor of two at very high energies. 
We also see a faster increase of the charm quark yield with the
colliding energy in AA collisions than that in pp collisions at high
energies. This is because the ratio of the charm quark yield in AA collisions
over that in pp collisions at the same energy is roughly the number of
binary collisions, which is proportional to the inelastic
nucleon-nucleon scattering cross section that increases with the colliding
energy.

We expect the string melting version of the AMPT
model~\cite{Lin:2001zk}  (instead of the default version of AMPT) to
be applicable in describing the dense matter at high energies when a
QGP is believed to be formed in the early stage of the collisions. 
Therefore we use the string melting version throughout this study. 
The hadronization of the partonic matter is accomplished by a 
spatial quark coalescence model~\cite{He:2017tla} after partons stop interacting. 
The open heavy flavor hadron species formed by quark coalescence
includes the following charm and bottom hadrons
at all possible charge states (plus the
corresponding anti-particles when applicable): $D$, $D^{*}$,
$D_{s}$, $D_{s}^{*}$, $\Lambda_{c}$, $\Sigma_{c}$, $\Xi_{c}$, $\Xi_{c}'$,
$\Xi_{cc}$, $\Omega_{c}$, $\Omega_{cc}$, $\Omega_{ccc}$, as well as $B$, $B^{*}$,
$B_{s}$, $B_{s}^{*}$, $B_{c}$, $B_{c}^{*}$, $\Lambda_{b}$, $\Sigma_{b}$,
$\Xi_{b}$, $\Xi_{b}'$, $\Xi_{bc}$, $\Xi_{bc}'$, $\Xi_{bb}$, $\Omega_{b}$,
$\Omega_{bc}$, $\Omega_{bc}'$, $\Omega_{bb}$, $\Omega_{bbc}$, and $\Omega_{bbb}$.
The hadron species are determined by the flavor combination of the two
or three coalescing (anti)quarks. In addition, for a pseudo-scalar
meson and a vector meson with the same flavor combination, our
previous approach is to form the meson to which the invariant mass of
the coalescing quark and antiquark is closer~\cite{Lin:2004en}. 
However, we find that the resultant vector to pseudo-scalar meson
ratios such as the $K^*/K$ ratio and the $D^*/D$ ratio are often far
away from the experimental data~\cite{Singha:2015fia}. 
Therefore, we change the previous approach and now set the ratio of
each type of vector to pseudo-scalar meson in the quark coalescence
model, 0.30 for primordial $\rho/\pi$, 0.50 for primordial $K^*/K$,
and 1.0 for primordial $D^*/D$ or $B^*/B$. 
For example, for all the flavor combinations that could form either
$D$ or $D^*$ mesons, we order them in terms of the excess mass 
(i.e. the difference between the two-quark invariant mass and the sum
of two quarks' masses) and assign the half with lower excess
masses to form $D$ and the half with higher excess masses to form
$D^*$. The above values are chosen to roughly reproduce the overall
magnitudes of the final vector to pseudo-scalar meson ratios observed
in $pp$ collisions of various energies. 
Note that the above values determine the ratios of the primordial
(i.e. right after quark coalescence) meson multiplicities in each
event, not the ratios in the final state that often include effects
from resonance decays and hadronic rescatterings. 

In the new quark coalescence model for AMPT~\cite{He:2017tla}, 
the overall relative probability of a quark to form a baryon instead
of a meson is determined by the $r_{BM}$ parameter, which is assumed
to be independent of the colliding energy and collision system. 
Generally, there would be no antibaryon formation if this parameter
is 0 but almost no meson formation if it goes to infinity.
In the updated AMPT model~\cite{Zhang:2019utb} used for this work, the
$r_{BM}$ value for light flavor $(u/d/s)$ hadrons is set to
0.53. On the other hand, the $r_{BM}$ value for heavy flavor hadrons
is set to 1.0, because using the light flavor value would lead to
too few charm baryons (by a factor of $\sim 4$) compared to the
experimental data in $pp$ or $AA$ collisions. 
More specifically, we determined this heavy flavor $r_{BM}$ value
according to the $\Lambda_c$ measurements in $pp$ collisions at LHC 
(as shown in Fig.~\ref{fig:pp_7000_D}b). 
In principle, the $r_{BM}$ value for charm hadrons is related to the
properties such as the number and masses of available charm baryon
states versus charm meson states, and the higher $r_{BM}$ value for
charm is consistent with the assumption that relative to light flavors
there are more charm baryon states than charm meson states
\cite{He:2019tik}.

Heavy quark productions directly depend on the PDFs as shown in
Eq.(\ref{eqn:dsigQQ}). 
In addition, for $AA$ collisions $f$ in
Eqs.(\ref{eqn:dsigQQ}-\ref{eqn:dsigjet})  
represents the parton PDFs in the nucleus instead of those in a free
nucleon and thus contains the nuclear shadowing functions. 
Therefore we expect that our recent updates to the AMPT
model~\cite{Zhang:2019utb} by including the newer parton distribution  
functions CTEQ6.1M~\cite{Stump:2003yu} of free nucleon and the modern
impact parameter-dependent EPS09s~\cite{Helenius:2012wd} nuclear 
shadowing functions should improve its descriptions of heavy flavor
productions. 

\section{Open charm results for $pp$ collisions}
\label{sec:charm_hadron_pp}

We now use the string melting version of the updated AMPT
model~\cite{Zhang:2019utb}, which already includes the heavy flavor
improvements  as detailed in the previous section, to study the open charm
production in $pp$ collisions in this section as well as $AA$
collisions in the next section.
Like the study on light flavors~\cite{Zhang:2019utb}, we set the Lund string 
fragmentation parameters to $a=0.8, b=0.4$ GeV$^{-2}$ for $pp$
collisions and $a=0.8, b=0.15$ GeV$^{-2}$ for the central Au+Au collisions at
RHIC energies and Pb+Pb collisions at LHC energies.
In addition, we set the parton elastic scattering cross section 
to $\sigma_p=3$ mb and the hadron cascade cutoff time to 30~fm$/c$.
The old AMPT results are often provided for comparisons, 
and there we use the AMPT version v2.26t9 with the same
parameters as in an earlier study~\cite{Lin:2014tya}.
Unless otherwise specified, the yield of each charm meson species
represents the average of the particles and the corresponding anti-particles.

\begin{figure}
\includegraphics[width=0.54\textwidth]{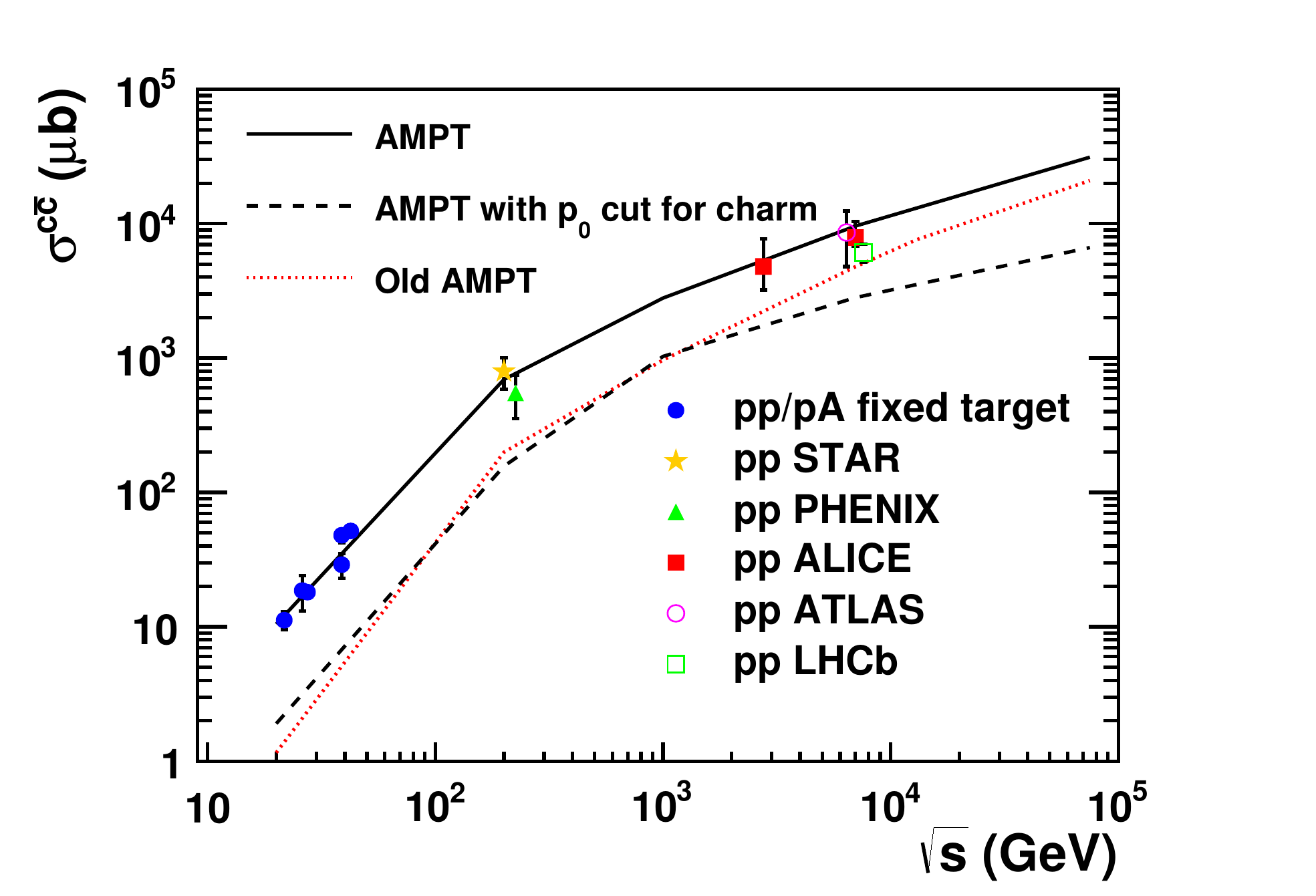}
\caption{(Color online) 
Total cross sections of charm-anticharm quark pairs from
  AMPT for $pp$ collisions in comparison with the world
  data~\cite{Lourenco:2006vw,Adare:2010de,Adamczyk:2012af,Aaij:2013mga,Aad:2015zix,Abelev:2012vra,Acharya:2017jgo}
  as functions of the colliding energy.
} 
\label{fig:sig_ccbar_energy_pp}
\end{figure}

The total $c\bar{c}$ pair cross section varying with $\sqrt{s}$ for $pp$
collisions from the AMPT model is shown in Fig.~\ref{fig:sig_ccbar_energy_pp}
compared with the available world data.
The data points from PHENIX, ATLAS and LHCb collaborations are
slightly shifted on horizontally. 
We see that the updated AMPT model 
with our recent modifications (solid line) can provide a
good description of the charm quark cross section in $pp$ collisions on
a wide energy range. 
On the other hand, the old AMPT model significantly under-estimates
the charm cross section, especially at low energies.
The dashed line shows the results when the charm quark production is subject
to the minimum $p_0$ requirement, where we see much lower charm quark
cross sections.

\begin{figure}[hbt]
\centering
\includegraphics[width=0.54\textwidth]{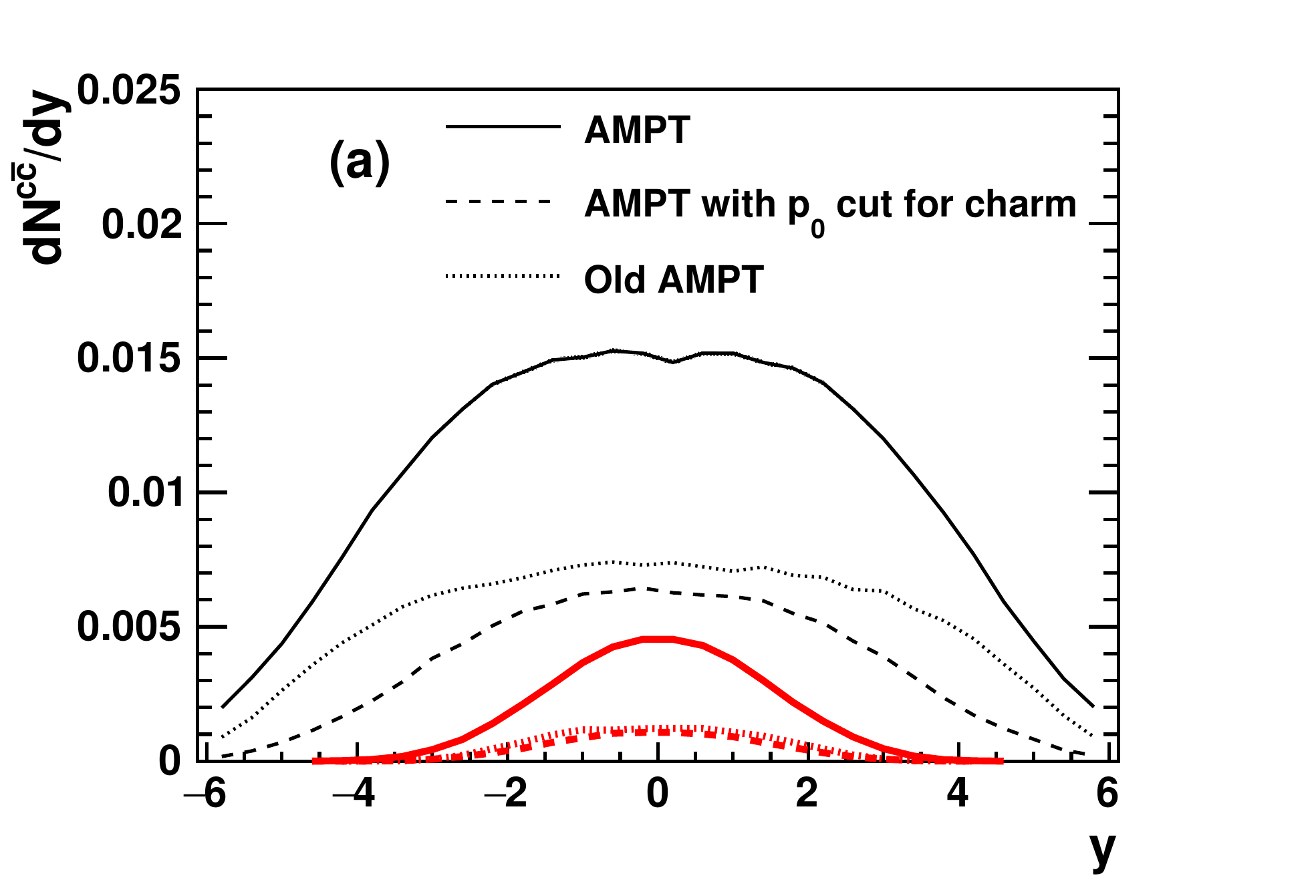}
\includegraphics[width=0.54\textwidth]{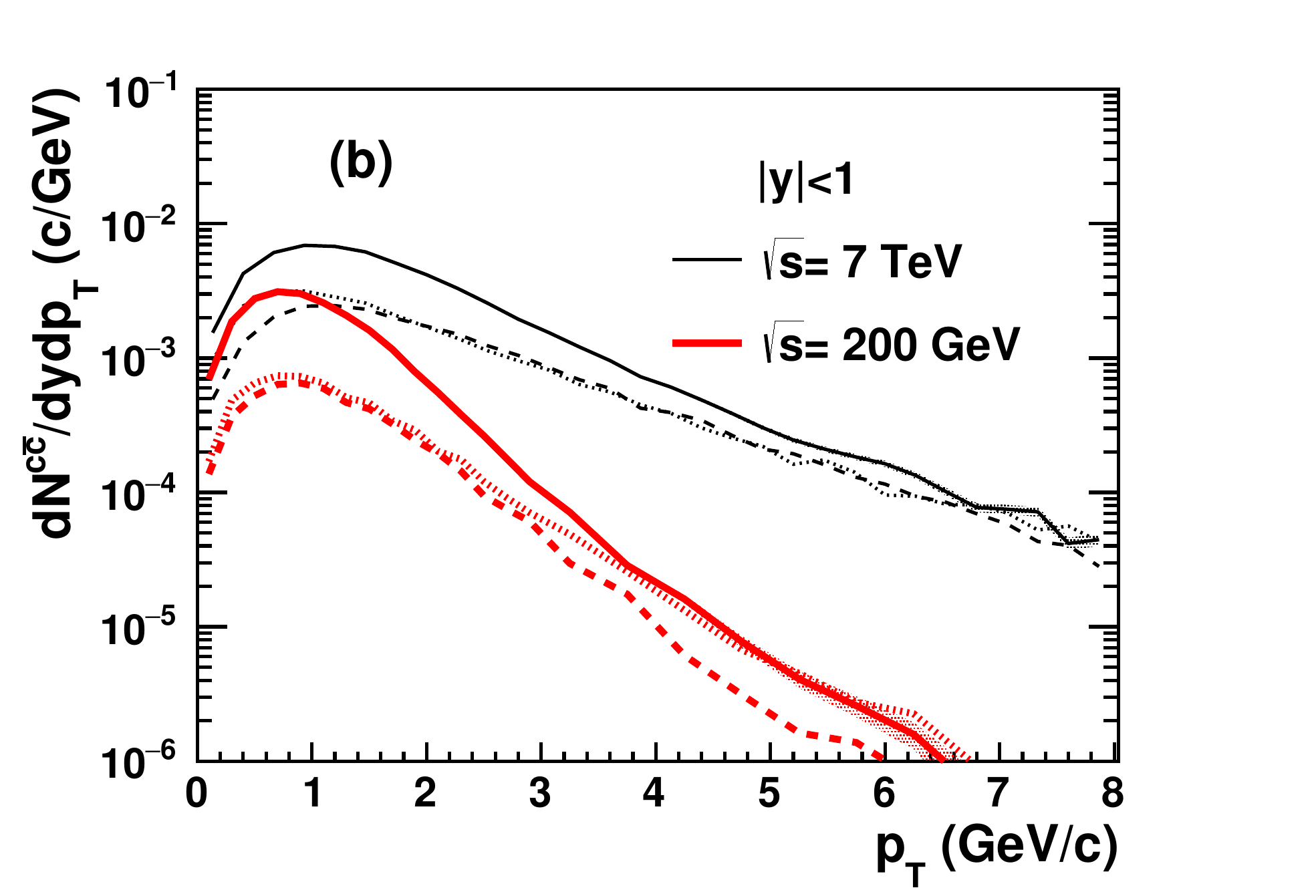}
\caption{(Color online) (a) Rapidity distributions and (b)
  transverse momentum spectra of charm quarks in $pp$
  collisions at $\sqrt{s}=200$ GeV (thick) and $\sqrt{s}=7$ TeV (thin)
  from the AMPT model. 
The shaded band represents statistical errors.
}
\label{fig:pp_cquark}
\end{figure}

We show the rapidity and transverse momentum distributions
of charm quarks from the AMPT model at $\sqrt{s}=200$ GeV and
$\sqrt{s}=7$ TeV in Fig.~\ref{fig:pp_cquark}.
We see the significant enhancement of the charm quark yield in the
updated AMPT compared to the old AMPT over all rapidities. 
The charm quark transverse momentum spectra at mid-rapidity are
shown in Fig.~\ref{fig:pp_cquark}(b), where
the statistical errors of the AMPT results are represented with a
shaded band. 
We also see that the removal of the transverse momentum cutoff $p_0$
for charm quarks mostly enhances the charm quark production in the low 
$p_{\rm T}$ region. It is interesting to observe in Fig.~\ref{fig:pp_cquark}
that the old AMPT result (dotted line) is rather similar to the result
from the updated AMPT model with the $p_0$ cut for charm quarks
(dashed curve); this is also seen in Fig.~\ref{fig:sig_ccbar_energy_pp}
until 1 TeV. 
Note that the old AMPT model also has a $p_0$ cut ($2$ GeV$/c$) on
the charm quark production. 

\begin{figure}[hbt]
\centering
\includegraphics[width=0.54\textwidth]{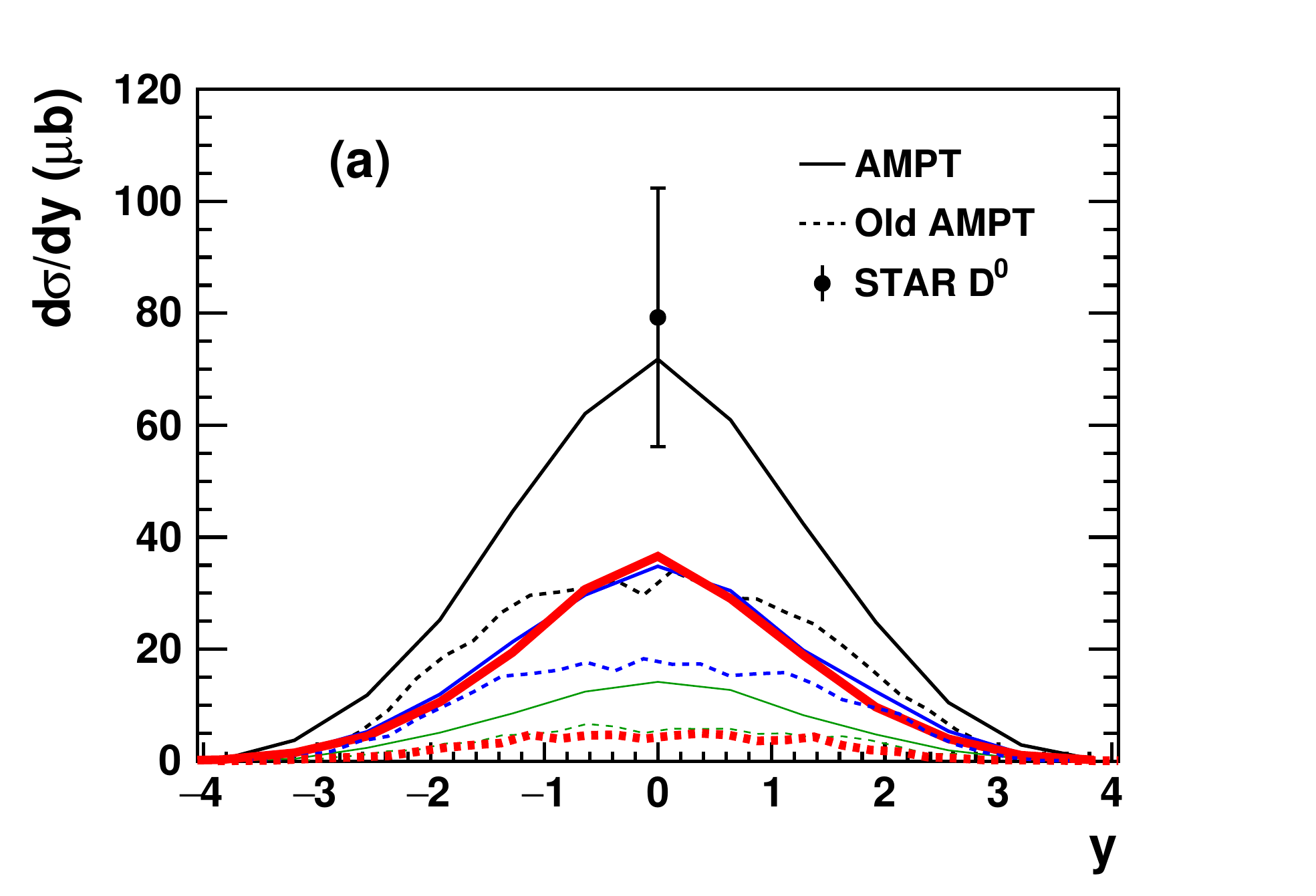}
\includegraphics[width=0.54\textwidth]{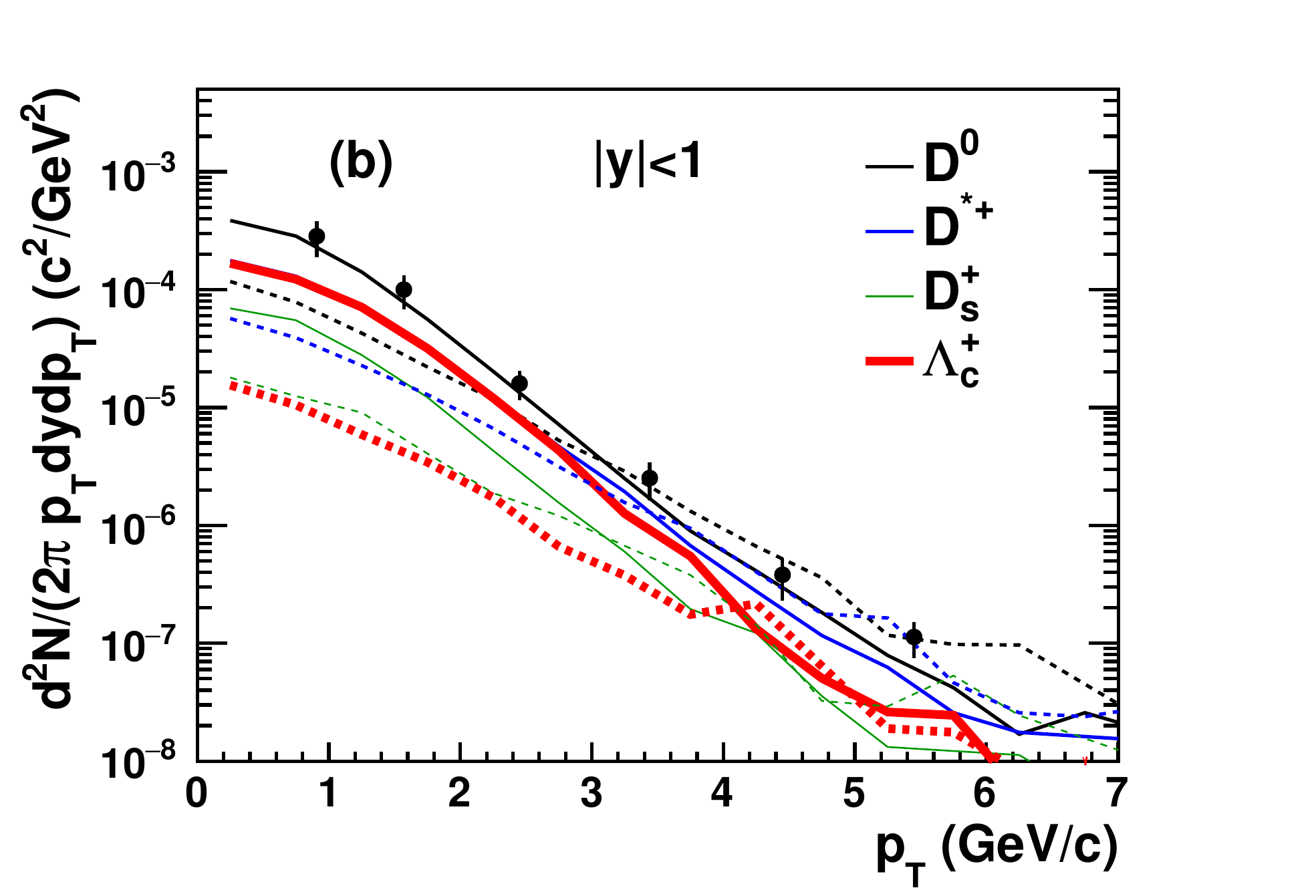}
\caption{(Color online) (a) Rapidity distributions and (b)
  transverse momentum spectra of open charm hadrons in $pp$
  collisions at $\sqrt{s}=200$ GeV from AMPT in comparison
  with the experimental data~\cite{Adamczyk:2012af,Adam:2018inb}.
}
\label{fig:pp_200_D}
\end{figure}

The productions of open charm hadrons in $pp$ collisions at
$\sqrt{s}=200$ GeV are shown in Fig.~\ref{fig:pp_200_D} for $D^{0}$,
$D^{*+}$, $D_{s}^{+}$ and $\Lambda_{c}^{+}$.
Note that the charm quark or hadron yields and spectra in this study
have been averaged over those for particles and their corresponding
anti-particles, e.g., the $\Lambda_{c}^{+}$ results
in Fig.~\ref{fig:pp_200_D} represent the average results
for $\Lambda_{c}$ and $\bar {\Lambda_{c}}$.
In addition, results of $D^0$ include both primordial $D^0$ mesons
and $D^0$ mesons from $D^*$ decays.
We see that the shape and magnitude of the charm hadron distributions
shown in Fig.~\ref{fig:pp_200_D} reflect those of the charm quarks
in Fig.~\ref{fig:pp_cquark}. For example, the $p_{\rm T}$ spectra
of charm hadrons from the updated AMPT model are much softer than those
in the old AMPT model, as also seen for the charm quarks.
The $D^{*}$ to $D^0$ ratio is rather flat across the shown $p_{\rm T}$ region.
We also see that results from the updated AMPT model is 
slightly lower than the STAR $D^0$ data but
still within the uncertainties.

\begin{figure}[hbt]
\centering
\includegraphics[width=0.47\textwidth]{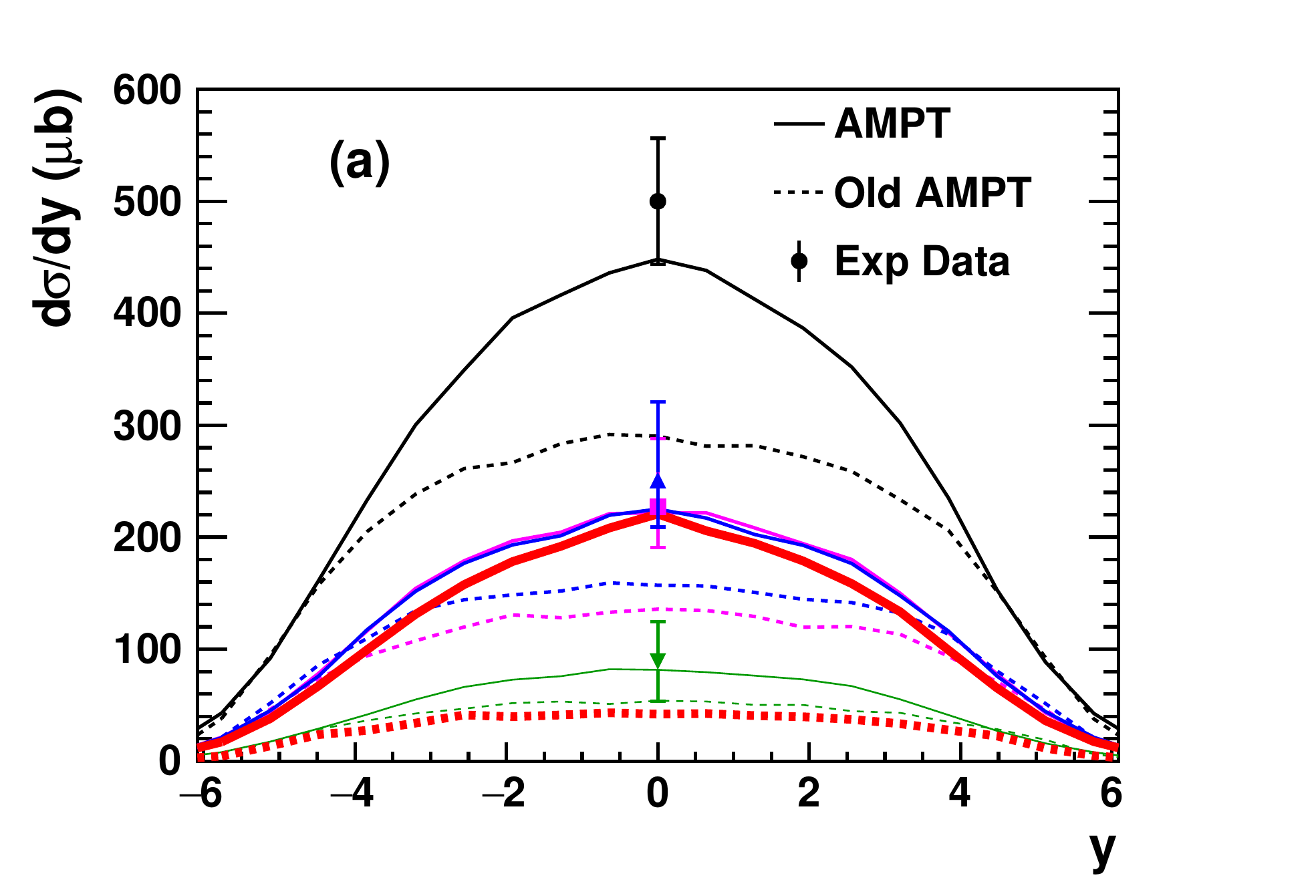}
\includegraphics[width=0.47\textwidth]{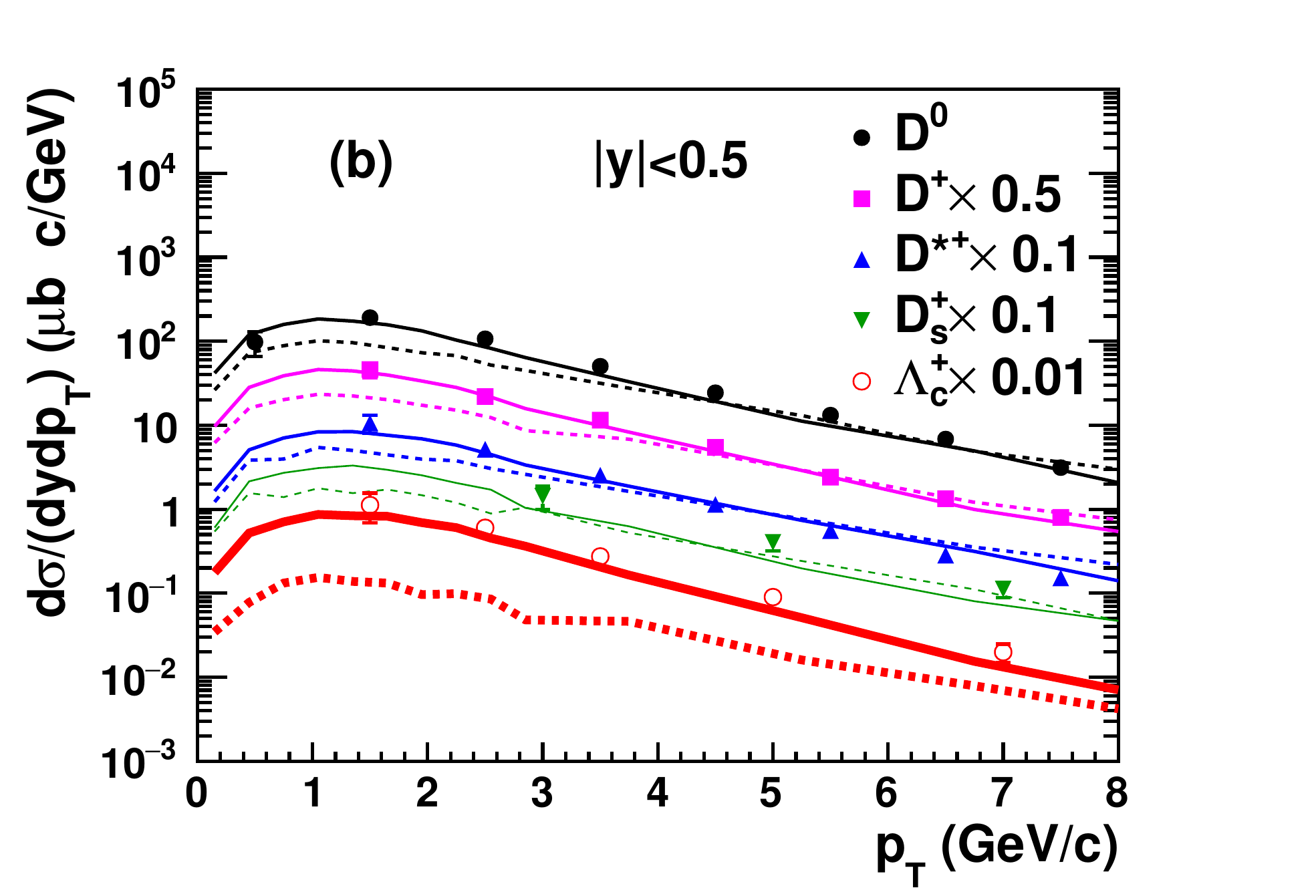}
\includegraphics[width=0.47\textwidth]{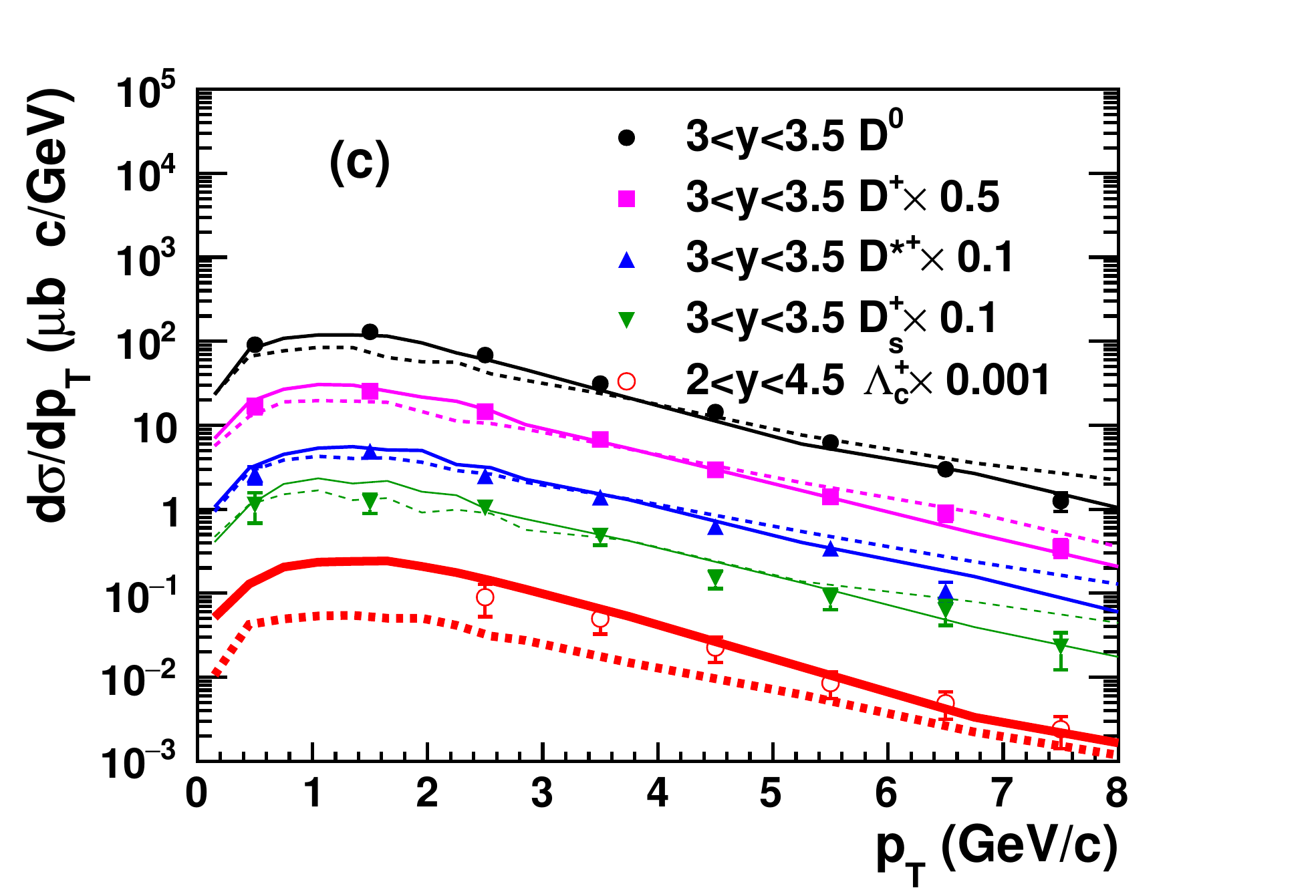}
\caption{(Color online) (a) Rapidity distributions and (b)
  mid-rapidity transverse momentum spectra of open charm hadrons in $pp$
  collisions at $\sqrt{s}=7$ TeV from AMPT in comparison with
  the ALICE $D$ meson data~\cite{Acharya:2017jgo} and $\Lambda_{c}$
  data~\cite{Acharya:2017kfy}. (c) Transverse momentum spectra of open charm hadrons in  $pp$ collisions at $\sqrt{s}=7$ TeV from AMPT in comparison with the LHCb data at forward rapidity~\cite{Aaij:2013mga}.
  }
\label{fig:pp_7000_D}
\end{figure}

In Fig.~\ref{fig:pp_7000_D} we confront the model results for
$D^0$, $D^+$, $D^{*+}$, $D_{s}^{+}$ and 
$\Lambda_{c}^{+}$ with the experimental data 
for $pp$ collisions at $\sqrt{s}=7$ TeV.
Fig.~\ref{fig:pp_7000_D}(a) shows that the results 
from the updated AMPT model for all these charm meson species 
roughly agree with the experimental data with large error bars,
which mainly come from the extrapolation to the unmeasured low $p_{\rm T}$
region in the experiment.
Note that the charm hadron yields for all species
in the updated AMPT model are much higher than those
in the old AMPT.
Figures~\ref{fig:pp_7000_D}(b) and Fig.~\ref{fig:pp_7000_D}(c) show
the transverse momentum distributions of these charm hadrons at
mid-rapidity and at forward rapidity, respectively, 
together with the experimental data. The updated AMPT model generally
describes the data for different charm meson species $D^0$, $D^+$, $D^{*+}$
and $D_{s}^{+}$.
The $\Lambda_{c}$ results are also roughly consistent with the
experimental data (after our tuning of the 
coalescence parameter $r_{BM}=1$ for heavy hadrons),  
including the $\Lambda_c$ spectrum at forward rapidity.

\section{Open charm results for $AA$ collisions}

\label{sec:charm_hadron_AA}

In $AA$ collisions, the charm production is subject to additional initial state
and final state effects. Initial state effects include the nuclear modification
to the parton distribution functions, while final state effects include parton
rescattering in ZPC parton cascade. We focus on central $AA$ collisions. Note
that for hadron level results, the event centrality class is determined
according to the charged particle multiplicity distribution generated by AMPT in
the same way as done by the STAR (within $|\eta|<0.5$)~\cite{Adam:2018inb} or
ALICE (within $-3.7<\eta<-1.7$ and $2.8<\eta<5.1$)
experiment~\cite{Acharya:2018hre}. For the parton level results, the event
centrality is obtained according to the impact parameter, where $b<4.88$ fm for
Au+Au collisions and $b<5.23$ fm for Pb+Pb collisions is used to sample the
0-10\% centrality events. We find that these two event centrality determination
techniques give very similar results in central $AA$ collisions such as the
0-10\% centrality.

\begin{figure}
\includegraphics[width=0.54\textwidth]{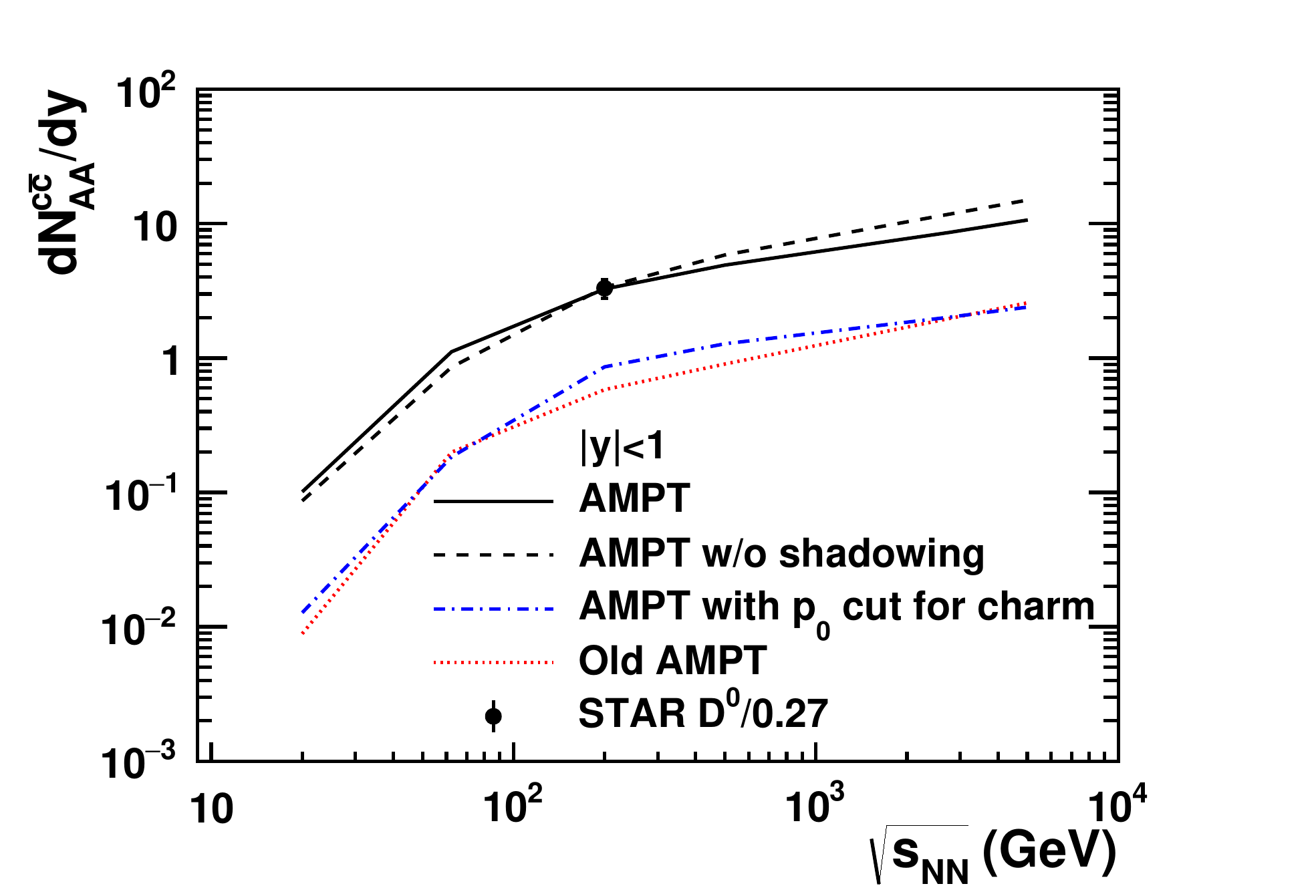}
\caption{Yields of charm quark pairs around mid-rapidity
 from AMPT for 0-10\% central Au+Au collisions at RHIC energies and
 Pb+Pb collisions above RHIC energies as functions of the colliding
 energy in comparison with the extracted STAR
 data~\cite{Adam:2018inb}. 
}
\label{fig:dndy_ccbar_energy_AA}
\end{figure}

In Fig.~\ref{fig:dndy_ccbar_energy_AA}, we show the charm quark yield at
mid-rapidity for 0-10\% central $AA$ collisions at different colliding energies.
We see that the EPS09s nuclear modification slightly enhances the charm yield at
low energies but significantly suppresses the charm yield at high energies
(above 1~TeV). This results from the anti-shadowing effect at large $x$ and the
shadowing effect at small $x$. 
We also see that the AMPT result agrees with the charm quark pair
yield for 0-10\% central Au+Au collisions at $\sqrt{s_{NN}}=200$ GeV
extracted from the STAR $D^0$ data~\cite{Adam:2018inb}.
The $D^{0}$ fragmentation fraction 0.27 is estimated based on
the STAR data of $d\sigma(D^{0})/dy=41$ $\mu b$ and
$d\sigma(D^{0}+D^{+}+D_{s}^{+}+\Lambda_c^+)/dy=152$ $\mu b$ for 10-40\% Au+Au
collisions at $\sqrt{s_{NN}}=200$ GeV~\cite{Xie:2018thr}. In contrast
to an earlier extraction~\cite{Adamczyk:2014uip} where the same $c
\rightarrow D^{0}$ fragmentation ratio (0.565) as in $p$+$p$ collisions was
assumed, the current $c \rightarrow D^{0}$ fraction takes into account
the $\Lambda_{c}$ and $D_{s}$ enhancements observed in the Au+Au
data. Compared to the old AMPT results (dotted line), the current AMPT 
results (solid line) for the charm quark yield at mid-rapidity are 
significantly higher (by a factor of $\sim$ 5). 
The main reason for this enhancement is our removal of the $p_0$ cut
for heavy flavor productions, as also shown by the dot-dashed
curve. In addition, the EPS09s nuclear shadowing implemented in the
updated AMPT model is generally weaker than the original parameterized
shadowing function in HIJING1.0.

\begin{figure}[hbt]
\centering
\includegraphics[width=0.54\textwidth]{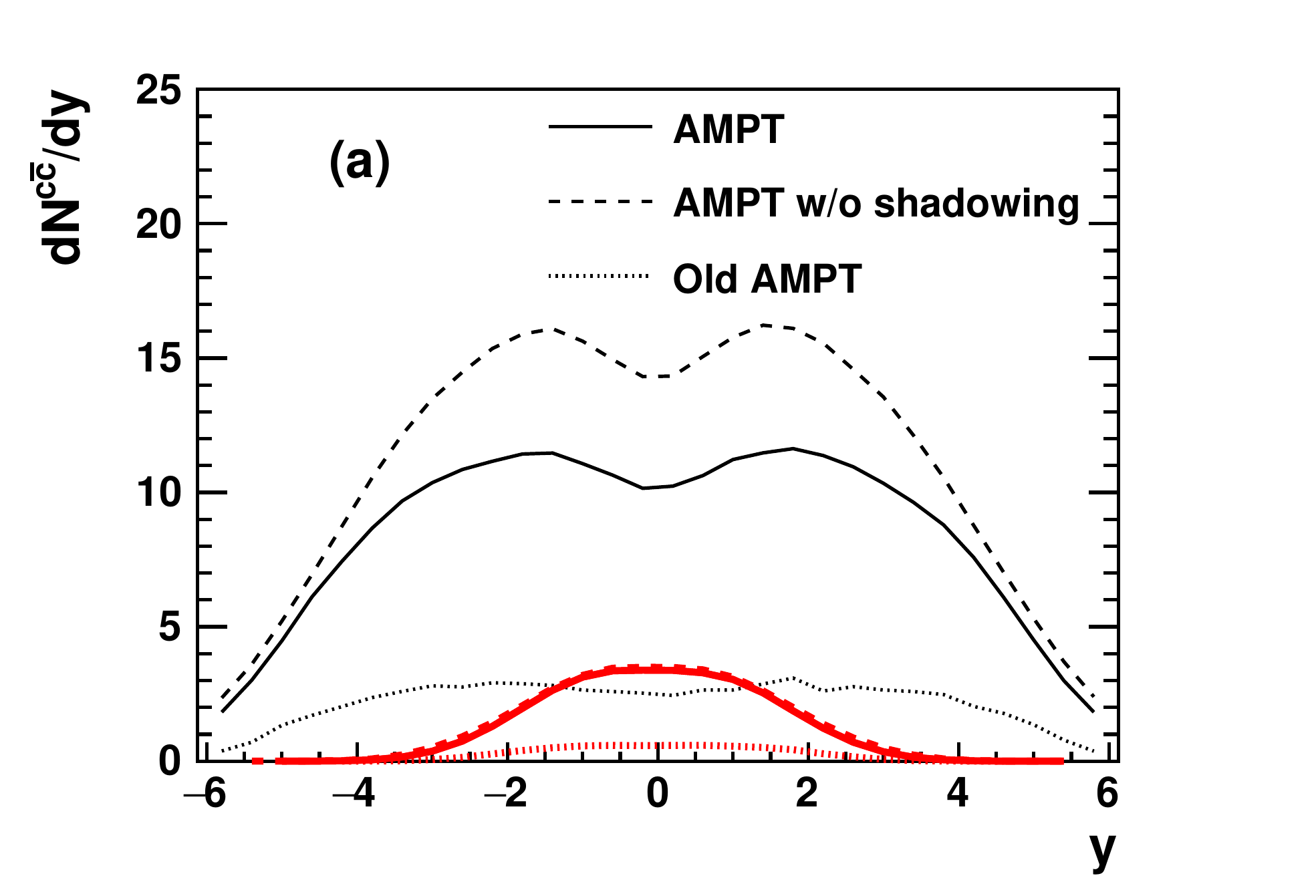}
\includegraphics[width=0.54\textwidth]{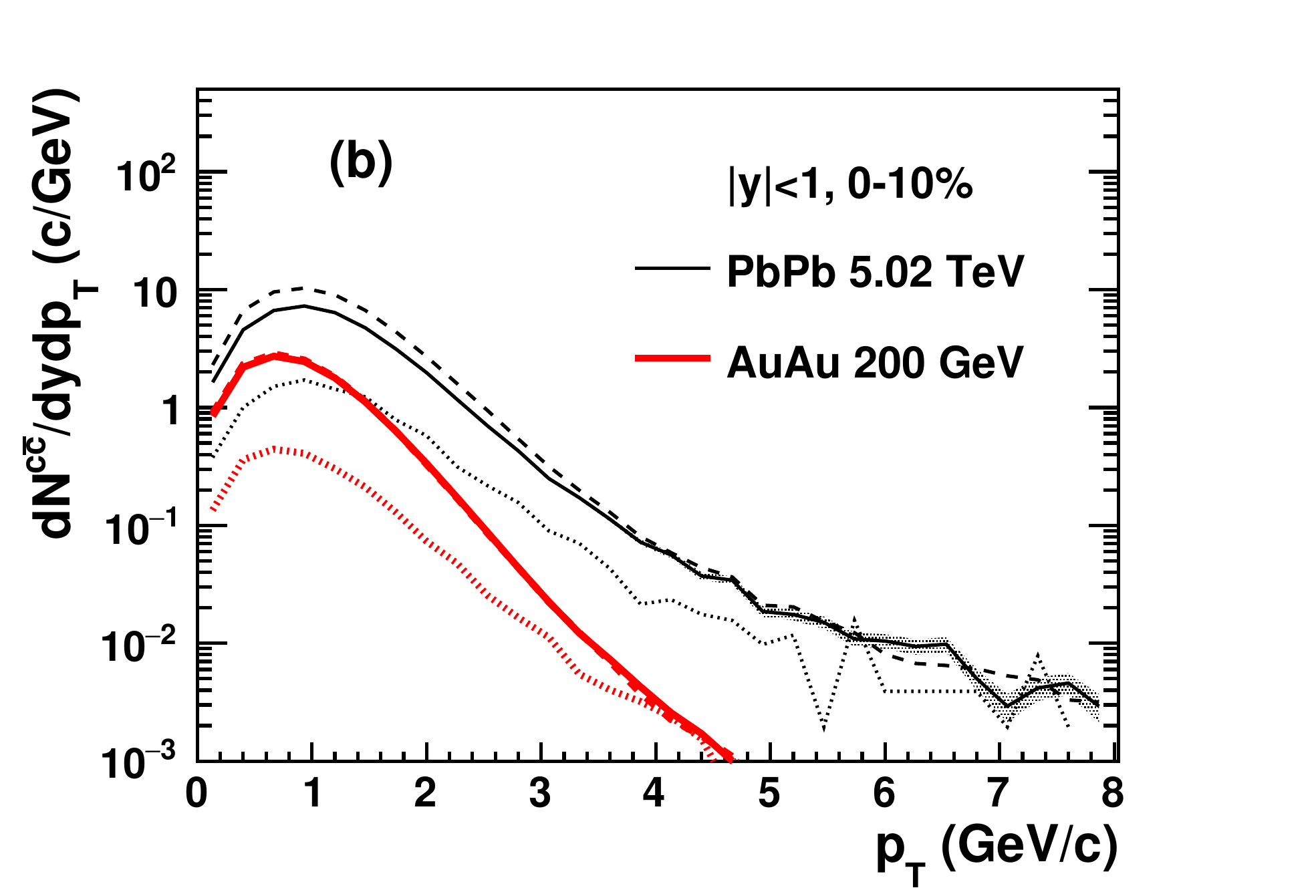}
\caption{(Color online) (a) Rapidity distributions and (b)
  transverse momentum spectra of charm quarks in central Au+Au
  collisions at $\sqrt{s_{NN}}=200$ GeV (thick) and central Pb+Pb
  collisions at $\sqrt{s_{NN}}=5.02$ TeV (thin) from the AMPT model.
}
\label{fig:AA_cquark}
\end{figure}

We show the charm quark distributions in
Fig.~\ref{fig:AA_cquark} for Au+Au collisions at
$\sqrt{s_{NN}}=200$ GeV and Pb+Pb
collisions at $\sqrt{s_{NN}}=2.76$ TeV for 0-10\% centrality.
We see in Fig.~\ref{fig:AA_cquark}(a) that 
the updated AMPT model (solid line) gives significantly more 
charm quarks compared to the old AMPT model (dotted line)
at both energies, which is already shown
in Fig.~\ref{fig:dndy_ccbar_energy_AA}.
On the other hand, nuclear shadowing leads to a $\sim$ 50\%
suppression of the rapidity density of charm quarks for central
Pb+Pb collisions at $\sqrt{s_{NN}}=5.02$ TeV.
We can find in Fig.~\ref{fig:AA_cquark}(b) that 
nuclear shadowing mainly suppresses the charm quark
yield in the low $p_{\rm T}$ region.
This is consistent with the fact that nuclear shadowing
is stronger at low $\mu_{F}^2$ that is associated with 
low $p_{\rm T}$ charm quarks.
Note that the EPS09s nuclear shadowing functions in the updated AMPT model
include the QCD evolution with the $\mu_{F}$, unlike the
shadowing parameterization implemented in HIJING1.0
as well as in the old AMPT model.

\begin{figure}[tbh]
\centering
\includegraphics[width=0.54\textwidth]{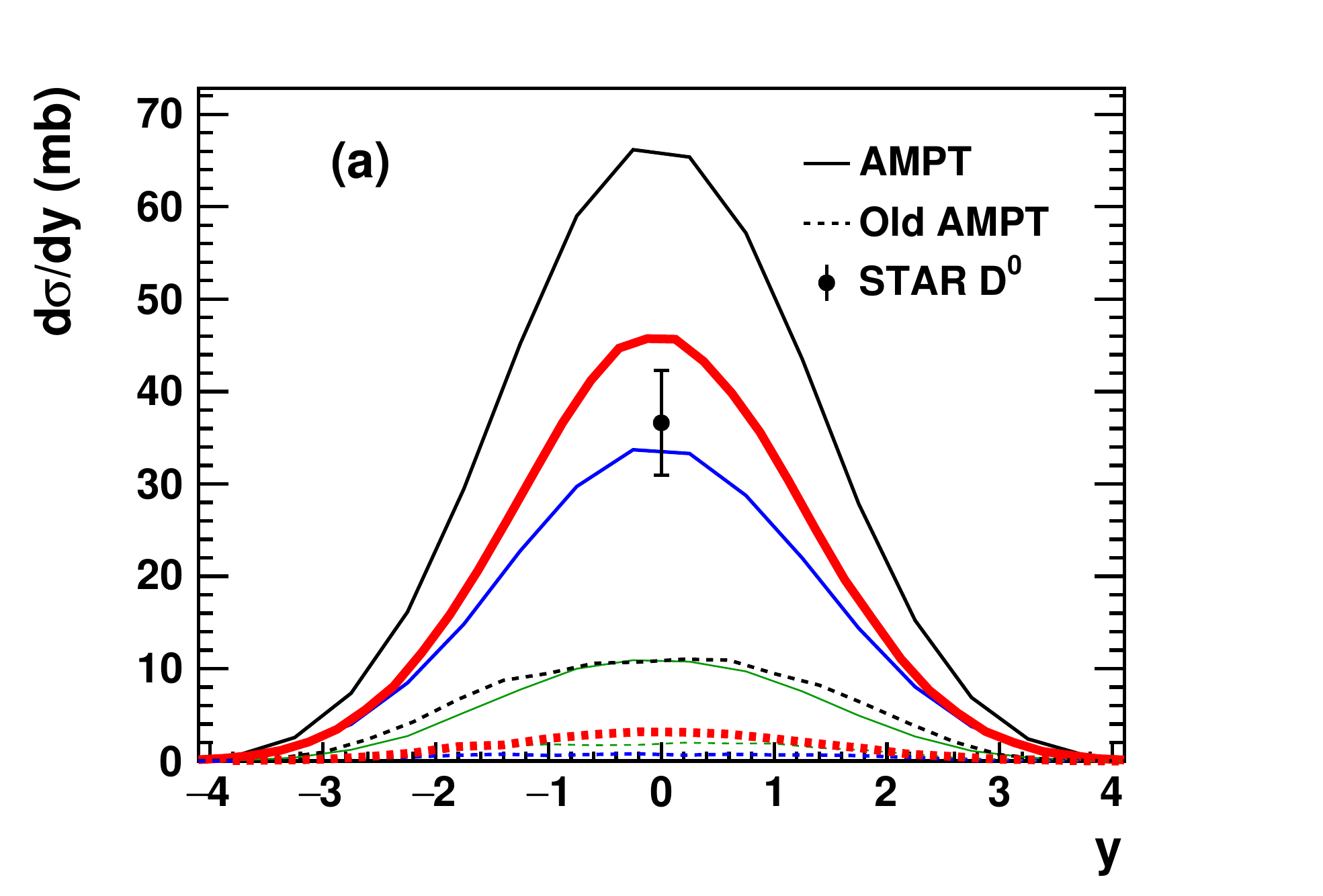}
\includegraphics[width=0.54\textwidth]{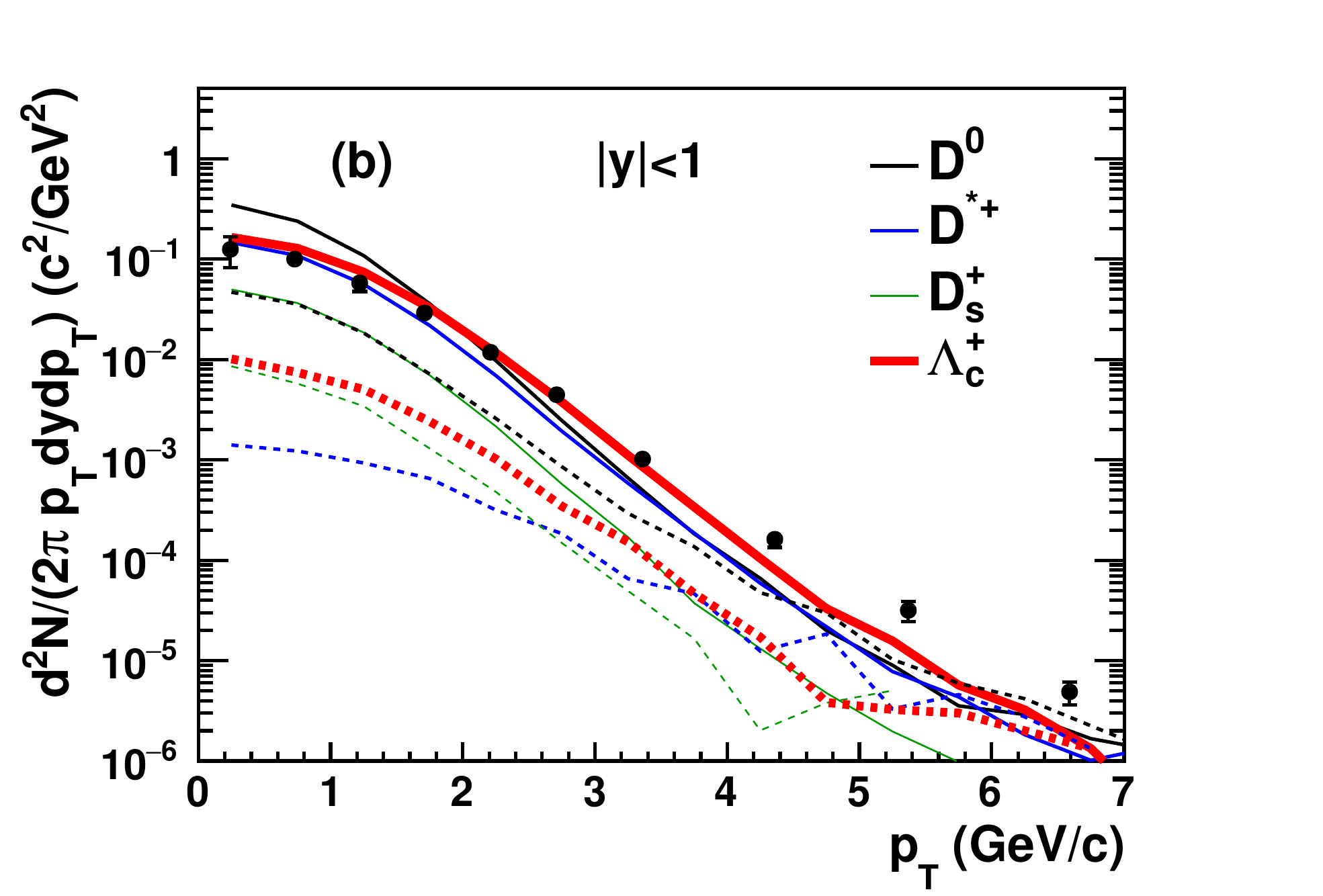}
\caption{ (Color online) (a) Rapidity distributions and (b)
  transverse momentum spectra of open charm hadrons
  in 0-10\% central Au+Au collisions at $\sqrt{s_{NN}}=200$ GeV
  from AMPT in comparison with the STAR $D^0$
  data~\cite{Adam:2018inb}. 
}
\label{fig:Au+Au_200_D}
\end{figure}

The productions of open charm hadrons including $D^0$, $D^{*+}$,
$D_s^+$ and $\Lambda_c$ in 0-10\% central Au+Au collisions 
at $\sqrt{s_{NN}}=200$ GeV are shown in Fig.~\ref{fig:Au+Au_200_D}.
As shown in Fig.~\ref{fig:Au+Au_200_D}(a), unlike the small 
$D^{*+}/D^0$ ratio in the old AMPT model,
the $D^0$ yield from the updated AMPT model is about twice
the $D^{*+}$ yield, close to the ratio observed in
Fig.~\ref{fig:pp_200_D}(a) for $pp$ collisions at the same energy.
Compared to the mid-rapidity STAR $D^{0}$ data, however, 
the AMPT result is significantly higher.
Also, the AMPT result on the $D^{0}$ $p_{\rm T}$
spectrum in Fig.~\ref{fig:Au+Au_200_D}(b) is too soft  
in comparison with the STAR data~\cite{Adamczyk:2014uip,Adam:2018inb}
and under-predicts the $D^{0}$ production at $p_{\rm T}>4$ GeV$/c$.
Since the yield of midrapidity charm quarks from AMPT is consistent
with the extracted STAR data as shown in Fig.~\ref{fig:dndy_ccbar_energy_AA}, 
the overestimation of the $D^0$ yield in AMPT could be because
the quark coalescence in AMPT gives fewer 
$D_s$ and $\Lambda_c$ than the data~\cite{Xie:2018thr}.
It has been suggested that a sequential coalescence of different charm
hadrons is important for the enhancement of $\Lambda_c/D^0$
and $D_s/D^0$ ratios in $AA$ collisions~\cite{Zhao:2018jlw}. Including
this sequential coalescence picture into AMPT could improve the descriptions
of different charm hadron species in the future.
In addition, the charm $p_{\rm T}$ spectra depend on the 
scatterings cross section and its angular distribution between charm
quarks and light flavors. The AMPT model currently uses the 
$g+g \rightarrow g+g$ cross section for scatterings between all parton
flavors, and improvements should be made to treat parton scatterings
between different flavors differently, 
where the comparison with the charm $p_{\rm T}$ spectra data will enable
us to extract the charm interaction strength with light flavors. 

\begin{figure}[hbt]
\centering
\includegraphics[width=0.54\textwidth]{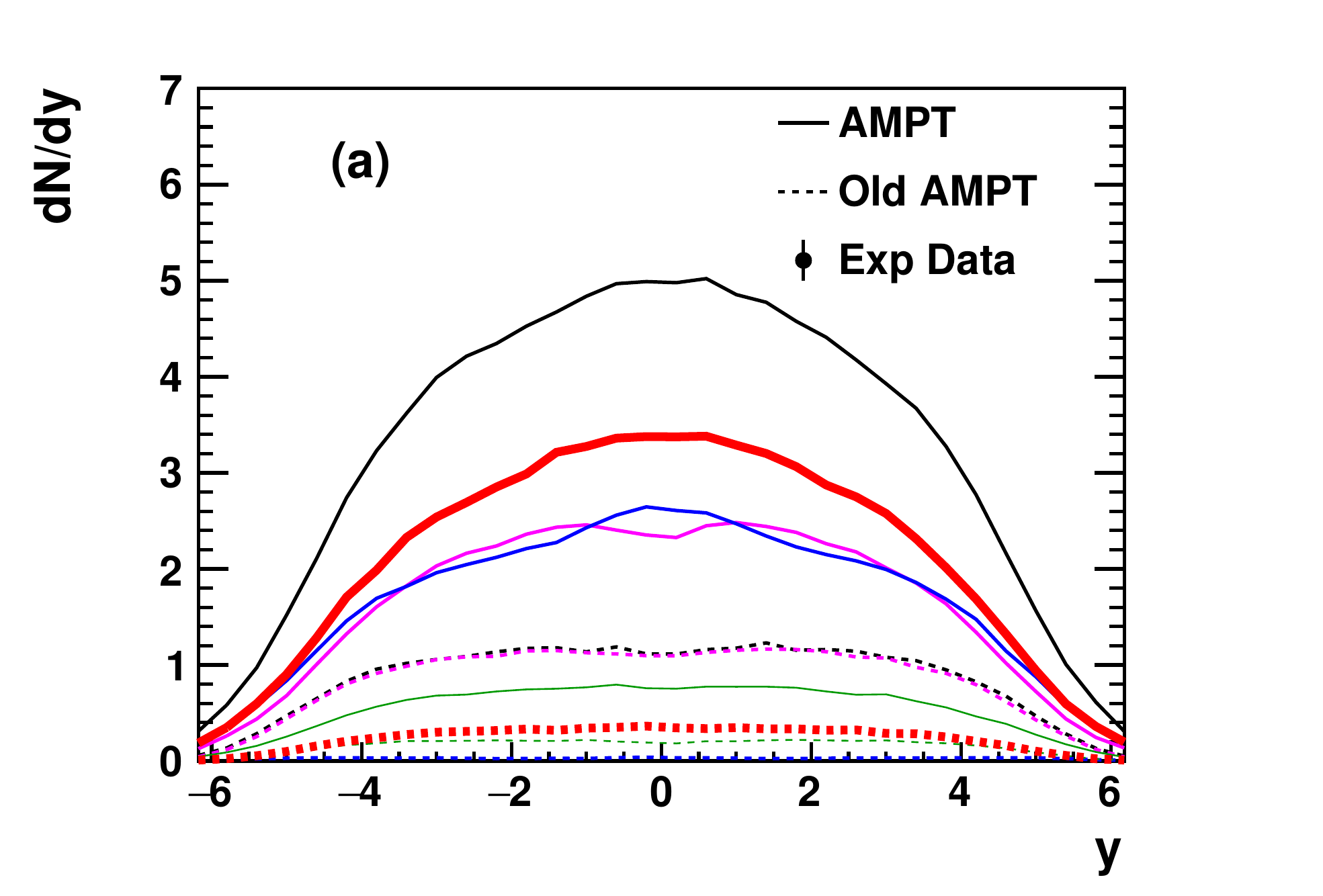}
\includegraphics[width=0.54\textwidth]{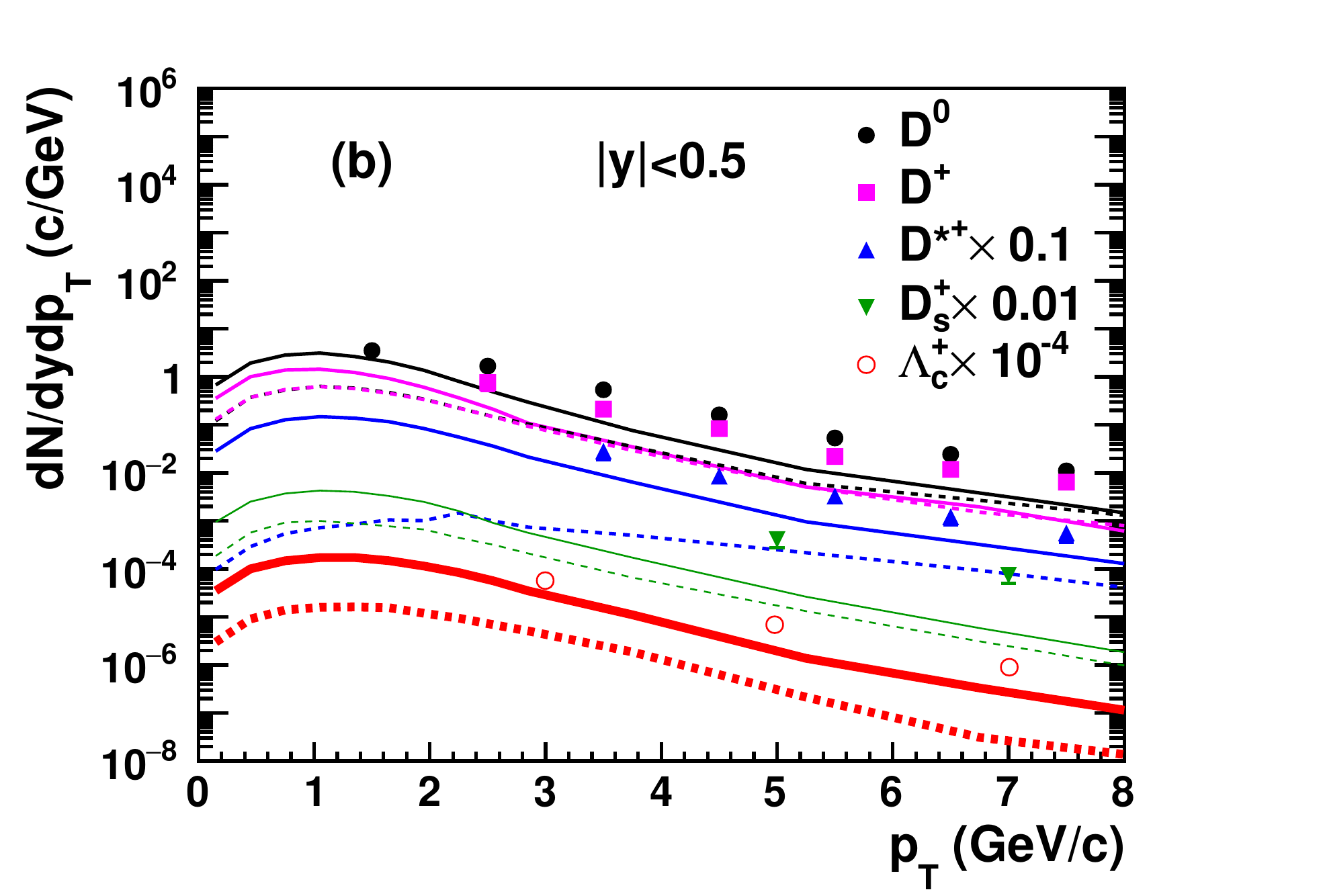}
\caption{(Color online) (a) Rapidity distributions and (b)
  transverse momentum spectra of open charm hadrons
  in 0-10\% central Pb+Pb collisions at $\sqrt{s_{NN}}=5.02$ TeV from AMPT
  in comparison with the ALICE data ~\cite{Acharya:2018hre,alice_Lambdac_sqm}. 
}
\label{fig:Pb+Pb_5020_D}
\end{figure}

\begin{figure*}[hbt]
\centering
\includegraphics[width=0.45\textwidth]{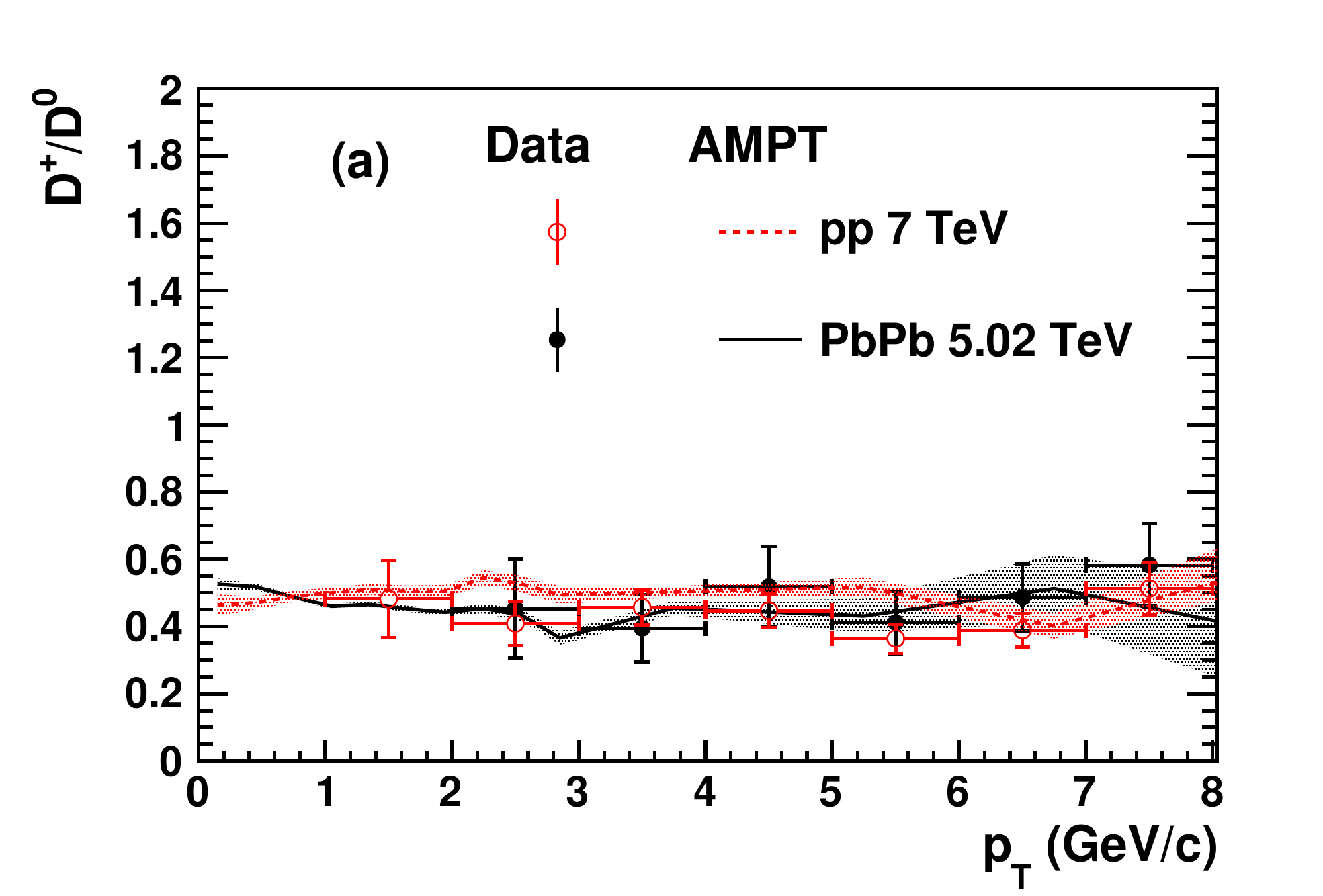}
\includegraphics[width=0.45\textwidth]{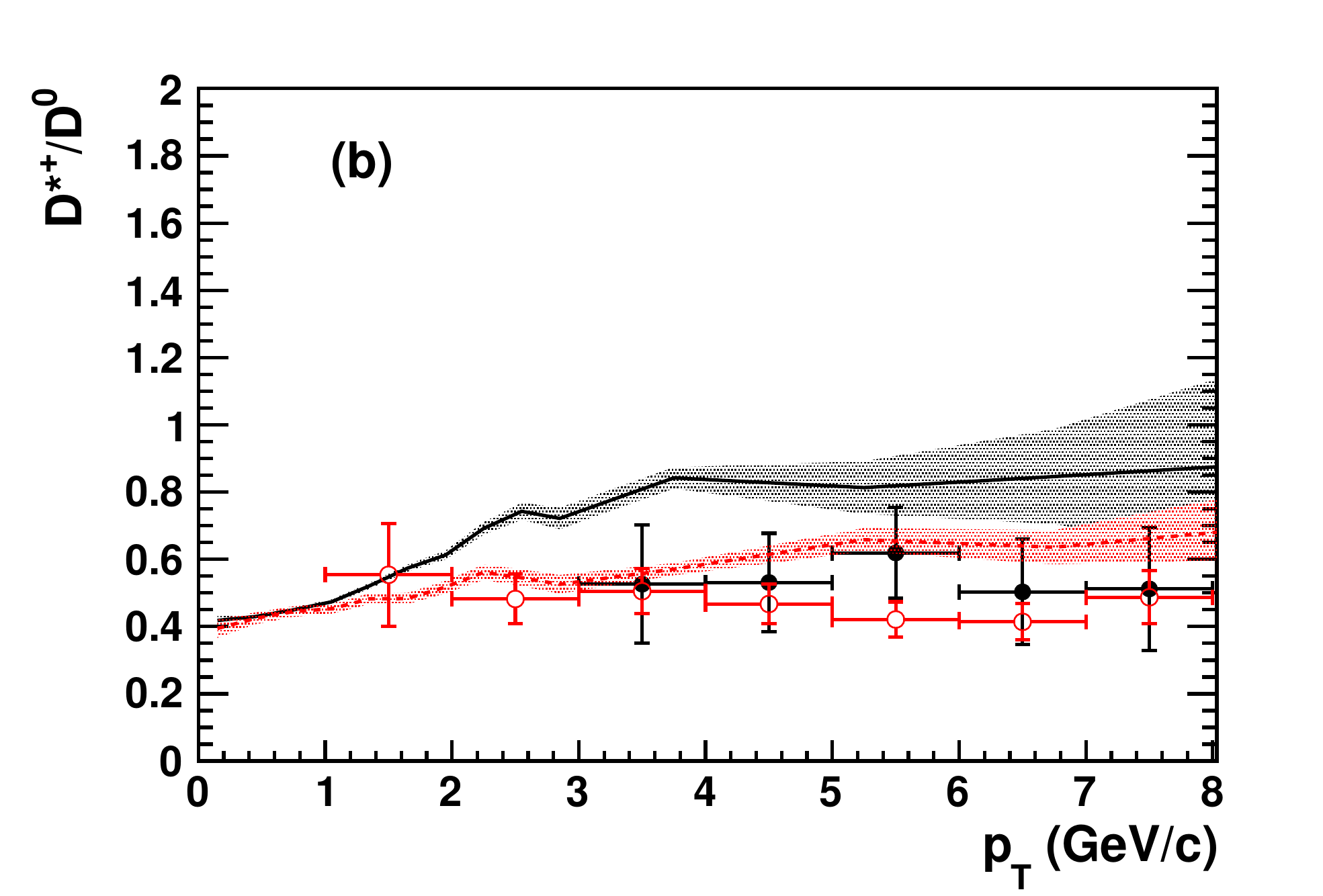}
\includegraphics[width=0.45\textwidth]{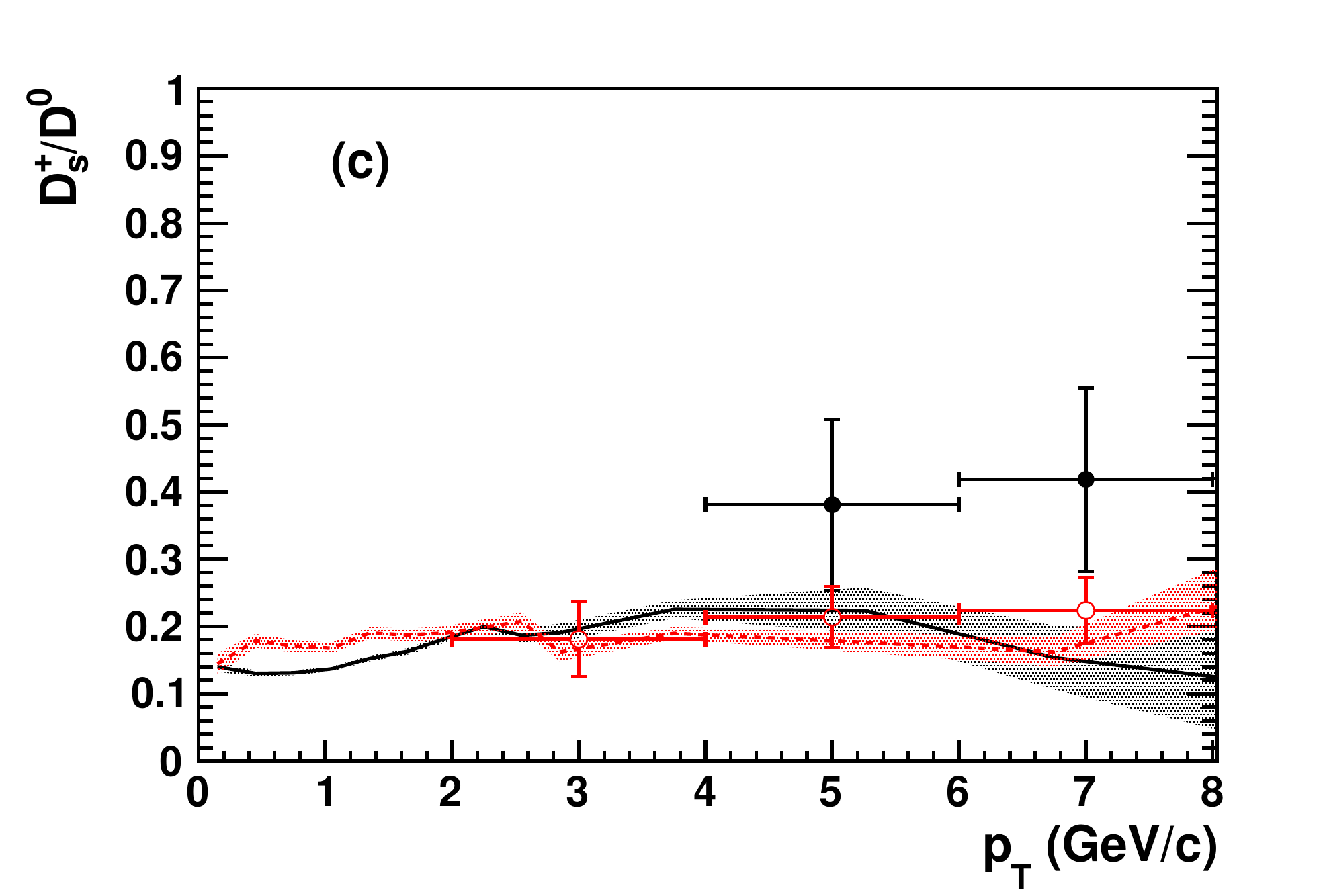}
\includegraphics[width=0.45\textwidth]{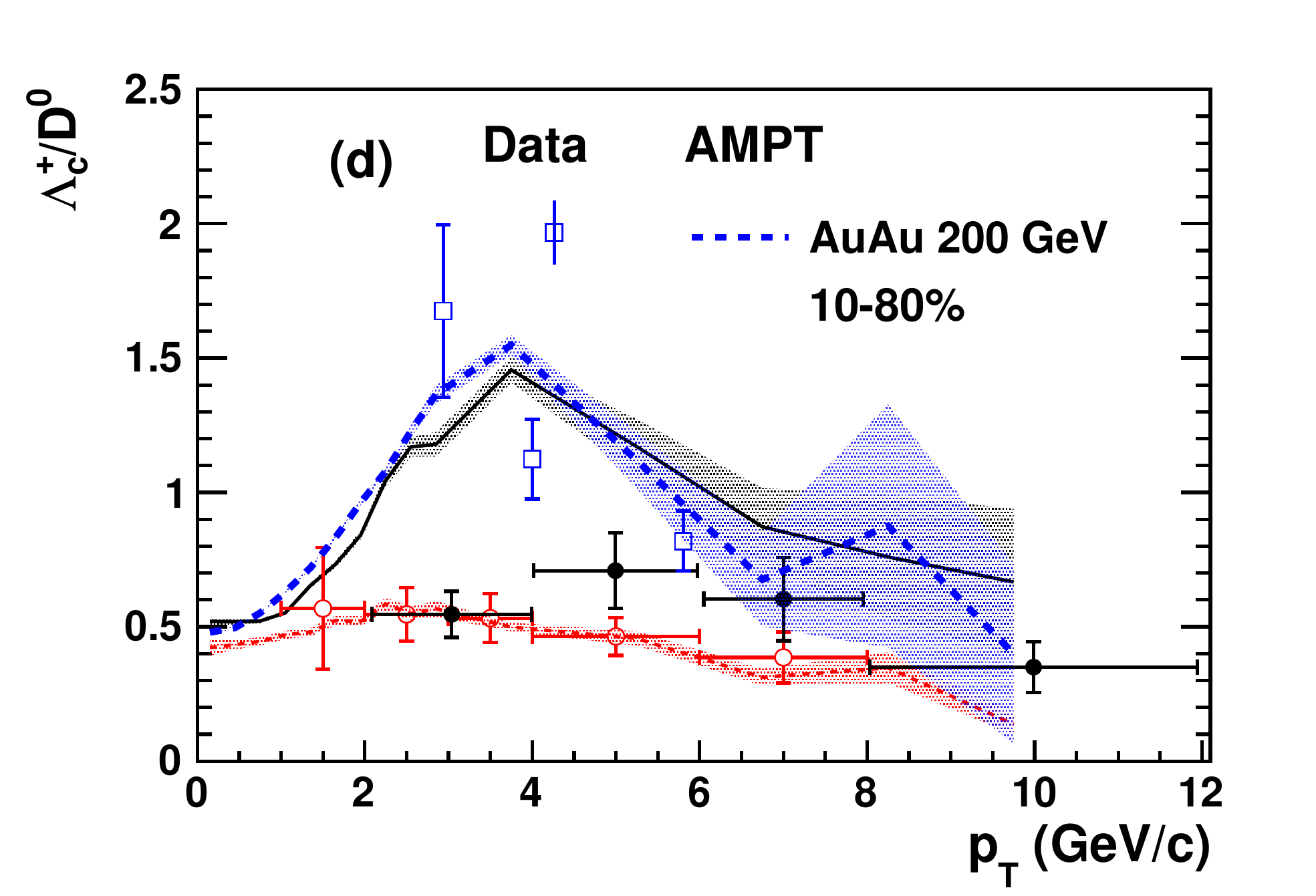}
\caption{(Color online) Ratios of mid-rapidity open charm hadron
  yields as functions of $p_{\rm T}$ in $pp$ collisions at
  $\sqrt{s}=7$ TeV and 0-10\% central Pb+Pb collisions at
  $\sqrt{s_{NN}}=5.02$ TeV from the AMPT model (curves)
  in comparison with the experimental data
\cite{Acharya:2017jgo,Acharya:2017kfy,Acharya:2018hre,alice_Lambdac_sqm}:
  (a) $D^{+}/D^0$, (b) $D^{*+}/D^0$, (c)
  $D^{+}_{s}/D^0$, and (d) $\Lambda_{c}^{+}/D^0$.
  Panel (d) also shows the AMPT result for 10-80\% central Au+Au
  collisions at 200 GeV in comparison with the STAR
  data~\cite{Xie:2018thr}.
} 
\label{fig:Dmeson_ratio_ALICE}
\end{figure*}

In Fig.~\ref{fig:Pb+Pb_5020_D} we show open charm hadrons in 0-10\%
central Pb+Pb collisions at $\sqrt{s_{NN}}=5.02$ TeV and compare with the ALICE
data. We first see that the model results for all these open charm particles are
lower than the experimental data, especially in the higher $p_{\rm T}$ region.
When we integrate the midrapidity $D^{0}$ yield at $p_{\rm T} \geq
1.0$ GeV$/c$, the ALICE data give 6.0 while the AMPT result gives
$3.3$. 
This underestimation of the $D^0$ yield in AMPT is first 
related to the branching of charm quarks into different hadron
species; e.g. the $\Lambda_c/D^0$ ratio from AMPT is higher than the
LHC data as shown in Fig.~\ref{fig:Dmeson_ratio_ALICE}(d).
Secondly, the $D^0$ $p_{\rm T}$ spectrum from the AMPT model is too
soft, as also seen at RHIC energies in
Fig.~\ref{fig:Au+Au_200_D}(b). 
Furthermore, the total charm quark yield in AMPT 
could be lower than that in the ALICE data, in part because of  
nuclear shadowing that has been shown in
Fig.~\ref{fig:dndy_ccbar_energy_AA} to 
significantly suppresses the charm yield at LHC
energies. Note that there is still a large uncertainty on the nuclear
shadowing of gluons~\cite{Helenius:2012wd}, which we have not explored
in this study.

We examine in Fig.~\ref{fig:Dmeson_ratio_ALICE} the ratios of open charm hadron
yields in 7 TeV $pp$ collisions and 0-10\% central Pb+Pb collisions at 5.02 TeV.
Note that, although the AMPT results of the charm hadron yields are
significantly lower than the experimental data for Pb+Pb collisions at
LHC, the ratios of different charm species are often compatible with
the data. It has been discussed that the $D_{s}^{+}$ production in
large collision systems may be enhanced due to the strangeness
enhancement, as the data in Fig.~\ref{fig:Dmeson_ratio_ALICE}(c) may
suggest. The AMPT results show no significant change in the
strange-to-nonstrange $D$ meson ratio in $AA$ collisions; 
however, sequential coalescence of different charm
species~\cite{Zhao:2018jlw} could change these results. 

Regarding the $\Lambda_{c}^{+}/D^0$ ratio shown in
Fig.~\ref{fig:Dmeson_ratio_ALICE}(d), the results from AMPT are
consistent with data in $pp$ collisions and show an enhancement in the
intermediate $p_{\rm T}$ region in $AA$ collisions. 
However, the current ALICE data
~\cite{Acharya:2017jgo,Acharya:2017kfy,Acharya:2018hre,alice_Lambdac_sqm} 
seem to favor a rather weak enhancement of charm baryons over charm
mesons. 
We also show the AMPT result for the 10-80\% centrality of Au+Au
collisions at 200 GeV, which shows a significant 
$\Lambda_{c}^{+}/D^0$ enhancement in rough agreement with 
the STAR data~\cite{Xie:2018thr}.

\section{Discussions}
\label{sec:discussions}

In the string melting AMPT model, the interactions
between charm quarks and the QGP medium are modeled
by parton elastic scatterings of ZPC. Thus our study
includes the collisional energy loss of heavy quarks but
neglects the radiative energy loss. Studies have suggested
that the elastic collisional energy loss is dominant for heavy
flavors below a moderately high $p_{\rm T}$, e.g., for charm
hadrons at $p_{\rm T}< \sim 5-6$ GeV$/c$ in Au+Au collisions at
$\sqrt{s_{NN}}$ = 200 GeV or at $p_{\rm T}< \sim 15$ GeV$/c$ in Pb+Pb collisions at 
$\sqrt{s_{NN}}$ = 2.76 TeV~\cite{Song:2015ykw}. Therefore results in our study
are applicable from low to moderately high $p_{\rm T}$
but not at very high $p_{\rm T}$.

In the ZPC parton cascade, the parton cross section and its angular distribution
determine the interaction strength between heavy quarks and the medium. Note
that any given scattering angular distribution can be exactly sampled with no
need of the assumption of small-angle scatterings; and this is an advantage of
the parton cascade approach. After the quark coalescence
process, the formed hadrons go through hadron interactions as modeled by an
extended ART model~\cite{Li:1995pra,Lin:2004en}. Currently we have not
implemented any hadron interactions for heavy hadrons except for decays of the
heavy hadron resonances.

It is also interesting to note that, while we have removed the $p_0$
cut for initial heavy flavor productions, 
we need to use a $p_0$ cut, which grows with the collision energy 
and the system size, for the initial light flavor minijet production 
to describe charged particle as well as light flavor multiplicities in
heavy ion collisions at high energies~\cite{Zhang:2019utb}. 
This different treatment of the $p_0$ cut for different flavors in the
final state seems to be inconsistent with initial state saturation models 
but might be understandable within final state saturation models 
such as the EKRT model~\cite{Eskola:1999fc}. 

\section{Summary}
\label{sec:summary}
In this work, we use the recently updated AMPT model to
study open heavy flavor productions.
In addition to the incorporation of modern parton distribution functions
in nuclei, we have removed the transverse momentum cutoff
$p_0$ for the pQCD heavy flavor production channels.
Systematic comparisons to the experimental data
show that the updated AMPT model can well describe the 
yields and $p_{\rm T}$ spectra of open charm hadrons
including $D$, $D^*$, $D_s$ and $\Lambda_c$
in $pp$ collisions at different energies. 
The updated model also describes the charm data
in central $AA$ collisions much better than the previous AMPT model,  
although it gives softer charm hadron spectra than 
the experimental data and also underestimates the open charm hadron
yields in central Pb+Pb collisions at LHC energies. 
These improvements in the AMPT model lay a 
foundation for further studies of heavy flavor observables 
together with light flavor observables 
within the transport model framework. 

\begin{acknowledgments}
We thank Xinye Peng for helpful discussions. This work is
supported by National Natural Science Foundation of China
under Grants No. 11890711, 11905188 and the
Fundamental Research Funds for the Central Universities, China
University of Geosciences (Wuhan) No. CUG180615.
\end{acknowledgments}

\bibliography{reference}

\end{document}